\documentclass[a4paper,fleqn]{cas-dc}

\usepackage[numbers]{natbib}
\usepackage{multicol}
\usepackage{amsfonts,amsmath}
\usepackage{textcomp}
\usepackage{stfloats}
\usepackage{url}
\usepackage{verbatim}
\usepackage{graphicx}
\usepackage{indentfirst}
\usepackage{psfrag,epsf}
\usepackage{enumerate}
\RequirePackage{amsthm}
\usepackage{amssymb}
\usepackage{booktabs}
\usepackage{bbm}
\usepackage{color}
\usepackage{bm}
\usepackage{siunitx}
\usepackage{subfigure}

\usepackage{makecell}
\usepackage{arydshln}
\usepackage{diagbox}
\usepackage{multirow}
\usepackage{array}
\usepackage{float}
\usepackage{cleveref}
\usepackage{algorithm}
\usepackage{algorithmic}
\usepackage{tikz}
\usepackage{threeparttable}
\usetikzlibrary{decorations.pathreplacing,calligraphy}
\usetikzlibrary{arrows,positioning,fit,shapes,calc}
\usepackage{dcolumn}%

\newcommand{\bee} {\begin{equation} }
\newcommand{\ene}{\end{equation}}

 \numberwithin{equation}{section}
 \numberwithin{lem}{section}

\DeclareGraphicsExtensions{.pdf,.png,.jpg}

\def\colred{\textcolor{red}}

\makeatletter%
\@addtoreset{equation}{section}

\newcommand\figcaption{\def\@captype{figure}\caption}
\newcommand\tabcaption{\def\@captype{table}\caption}
\makeatother%

\def\tsc#1{\csdef{#1}{\textsc{\lowercase{#1}}\xspace}}
\tsc{WGM}
\tsc{QE}
\tsc{EP}
\tsc{PMS}
\tsc{BEC}
\tsc{DE}

\begin{document}
\let\WriteBookmarks\relax
\def\floatpagepagefraction{1}
\def\textpagefraction{.001}
\let\printorcid\relax
\shorttitle{Robust Outlier Detection and Low-Latency Concept Drift Adaptation for Data Stream Regression}
\shortauthors{Bingbing Wang et~al.}

\title [mode = title]{Robust Outlier Detection and Low-Latency Concept Drift Adaptation for Data Stream Regression: A Dual-Channel Architecture}          

\author[school1]{Bingbing Wang} 
\fnmark[1]
\ead{bbwangstat1@stu.suda.edu.cn}
\author[school2]{Shengyan Sun} 
\fnmark[1]
\ead{2262403013@stu.suda.edu.cn}
\author[school1]{Jiaqi Wang} 
\ead{20234007011@stu.suda.edu.cn}
\author[school2]{Yu Tang} 
\ead{ytang@suda.edu.cn}
\corref{corresponding}

\cortext[corresponding]{Corresponding author.}
\fntext[equal]{Bingbing Wang and Shengyan Sun have contributed equally to this work.}

\affiliation[school1]{organization={School of Mathematical Sciences},
            addressline={Soochow University},
            city={Suzhou},
            postcode={215031},
            state={Jiangsu},
            country={China}}
            
\affiliation[school2]{organization={School of Future Science and Engineering},
            addressline={Soochow University}, 
            city={Suzhou},
            postcode={215006}, 
            state={Jiangsu},
            country={China}}
            
\begin{abstract}
Outlier detection and concept drift detection represent two challenges in data analysis. Most studies address these issues separately. However, joint detection mechanisms in regression remain underexplored, where the continuous nature of output spaces makes distinguishing drifts from outliers inherently challenging. To address this, we propose a novel robust regression framework for joint outlier and concept drift detection. Specifically, we introduce a dual-channel decision process that orchestrates prediction residuals into two coupled logic flows: a rapid response channel for filtering point outliers and a deep analysis channel for diagnosing drifts. We further develop the Exponentially Weighted Moving Absolute Deviation with Distinguishable Types (EWMAD-DT) detector to autonomously differentiate between abrupt and incremental drifts via dynamic thresholding. Comprehensive experiments on both synthetic and real-world datasets demonstrate that our unified framework, enhanced by EWMAD-DT, exhibits superior detection performance even when point outliers and concept drifts coexist.
\end{abstract}


\maketitle

\section{Introduction}
\label{section:1}
In the era of big data and machine learning, the paradigm of data analytics has fundamentally shifted from static, batch-based processing to real-time stream analytics. From industrial sensor monitoring to financial market forecasting, data streams arrive in a continuous and sequential fashion \cite{wares2019data}. Unlike offline learning, where data is assumed to be stationary and identically distributed, real-world data streams are characterized by their intrinsic volatility and non-stationarity. Consequently, the ability to maintain predictive accuracy over time has become the core goal of deploying machine learning models in dynamic environments.

\begin{figure*}[b]
\centering
\scalebox{0.5}{
\includegraphics{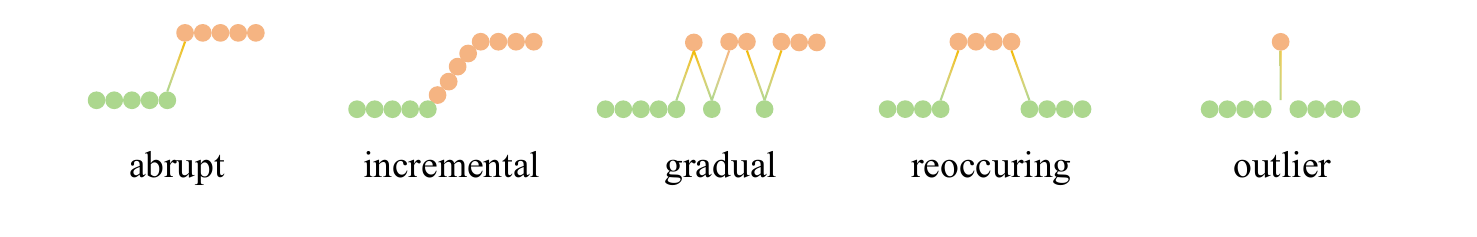}
}
\caption{Patterns of changes over time (outlier is not concept drift).}
\label{fig:drift_type}
\end{figure*}

The primary obstacle in stream learning is concept drift—the phenomenon where the joint distribution of input and target variables changes over time. If a model fails to adapt to these changes, its performance will inevitably degrade, a failure mode that is unacceptable in safety-critical applications. While existing literature has extensively categorized drifts into abrupt, incremental, gradual and recurring patterns (as illustrated in Fig. \ref{fig:drift_type}), accurately detecting them in real time remains a formidable challenge. This challenge is exacerbated in regression tasks, where the output space is continuous and infinite, making the definition of ``decision boundaries'' more ambiguous compared to classification tasks.

However, the challenge of concept drift does not exist in a vacuum. A more insidious problem, often overlooked in idealized experimental settings, is the pervasive presence of outliers—anomalies caused by sensor failures, transmission errors, or malicious attacks \cite{sullivan2021so}. The co-occurrence of outliers and concept drift creates a ``Chicken-and-Egg'' dilemma for regression models, leading to two fatal failure modes. First, traditional regression learners suffer from the masking effect, where extreme outliers skew model parameters, thereby concealing genuine distributional drifts. Second, the false alarm problem arises when sequences of outliers are misinterpreted as abrupt drifts, triggering unnecessary model resets that cause catastrophic forgetting of valuable historical knowledge. Despite the urgency of this issue, the literature remains surprisingly fragmented. Extensive research has been dedicated to Outlier Detection (OD) and Concept Drift Detection (CDD) separately (see \cite{kiani2024survey} \cite{lima2022learning} and reference therein). Furthermore, the vast majority of detection frameworks focus on classification and clustering tasks (see \cite{palli2024online} \cite{gemaque2020overview} and reference therein). In the realm of regression, a significant void remains for joint detection mechanisms, stemming from the fact that defining the confidence interval for residuals is structurally more complex than tracking discrete misclassification rates \cite{laory2013combined}.

\begin{figure*}
\centering
\scalebox{0.5}{
\includegraphics{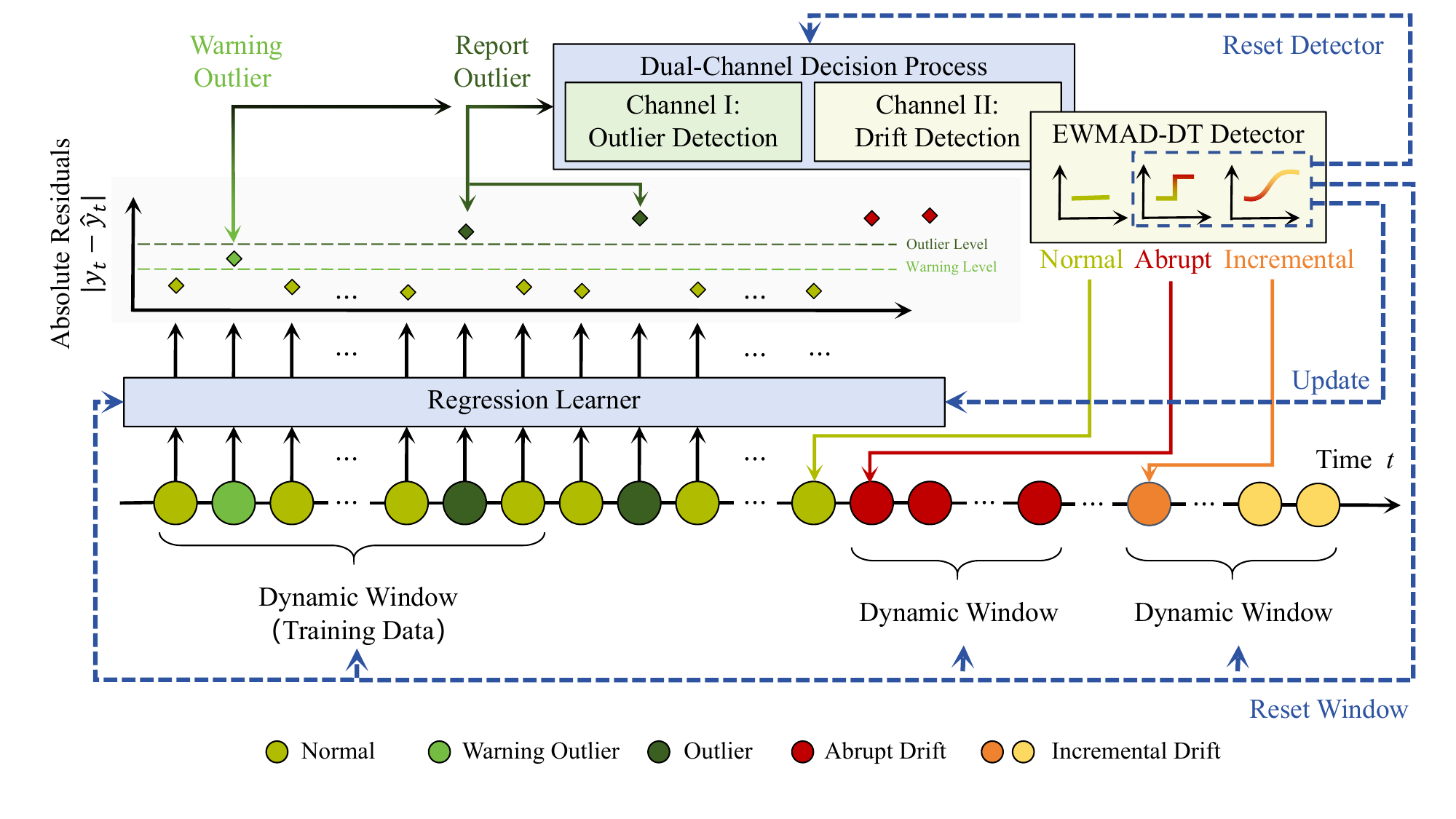}
}
\caption{Architecture of joint outlier and concept drift detection on labeled data streams. Initially, data points within a dynamic window are used to train a regression learner. Subsequently, the absolute residual of each data point is combined with the Dual-Channel Decision Process to detect whether the point is a warning outlier, outlier, abrupt drift or incremental drift. Once a concept drift is detected, the window and EWMAD-DT detector will be reset, and the learner will be updated accordingly.}
\label{fig2:env}
\end{figure*}

To overcome this challenge, we argue that OD and CDD must be treated as a dual-channel decision process: outliers must be filtered to protect the drift detector, while drift must be confirmed to update the outlier threshold.
In this paper, we propose a novel joint regression framework capable of simultaneously detecting outliers and concept drifts in evolving data streams. Unlike traditional approaches that treat data irregularities homogeneously, our framework leverages robust regression learners to minimize the impact of noise on parameter estimation. Building on this robust foundation, we design a dual-channel mechanism that explicitly disentangles point outliers from genuine distributional shifts. We focus specifically on abrupt and incremental drifts as they represent the two extremes of the conceptual change spectrum—instantaneous shifts versus continuous evolutions. By mastering these two distinct prototypes alongside point outliers, our framework establishes a foundational baseline for robustness, where other complex forms such as gradual and reoccurring drifts are implicitly covered as intermediate states or sequential abrupt events. Consequently, our system can not only flag the occurrence of a change but also diagnose its type—effectively distinguishing between outliers, abrupt drifts, and incremental drifts. Our framework is shown in Fig. \ref{fig2:env}.

The main contributions of this paper are summarized as follows:
\begin{itemize}
    \item A Model-Agnostic Joint Detection Framework: We propose a novel dual-channel decision process that unifies outlier and concept drift detection for stream regression. By breaking the dependency between data cleaning and model updates, this framework serves as a universal solution, capable of equipping any robust regression model with the ability to adapt to both concept drifts and outliers.
    \item Fine-Grained Drift Detector: We propose EWMAD-DT, a novel detector that leverages dynamic thresholds to differentiate between abrupt and incremental drift. By monitoring the trend of robust prediction residuals, it triggers appropriate adaptation strategies based on the specific type of change detected.
    \item Robust Experimental Results: Extensive experiments, utilizing both synthetic and real-world datasets comprehensively evaluate the performance of our framework. The results demonstrate that our robust framework, empowered by EWMAD-DT, achieves superior predictive performance and remarkable detection accuracy in scenarios characterized by severe concept drift and outlier contamination.
\end{itemize}

The remaining sections of this paper are organized as follows. In Section \ref{section:Related work}, we provide an overview of the latest research relevant to our study. In Section \ref{section:Methodology}, we introduce a detection framework that employs a dual-channel decision process for prediction residuals. In Section \ref{section:Experimental evaluation}, we evaluate the proposed method on both synthetic and real-world datasets. Finally, in Section \ref{section:Conclusion}, we conclude the paper with a discussion of our findings and contributions. 

\section{Related work}
\label{section:Related work}
\subsection{Outlier detection methods}
\label{covariate drift detection}
In regression tasks, classic leave-one-out methods based on residual analysis, such as Cook's distance and DFFITS, are suitable for detecting an individual influential point \cite{cousineau2010outliers}. 
To address these limitations, robust regression techniques were developed to provide reliable estimates even under high contamination. 
Specifically, Huber Regression modified the standard least-squares objective by employing a hybrid loss function—quadratic for small residuals and linear for large ones—thereby down-weighting outliers without completely discarding information \cite{jadon2024comprehensive}.
In a non-parametric vein, the Theil-Sen Estimator computed the median of slopes derived from all pairs of sample points, offering strong resilience against outliers in both the response and predictor variables \cite{raymaekers2021fast}.
Meanwhile, Random Sample Consensus (RANSAC) adopted a stochastic resampling strategy, iteratively fitting models to minimal data subsets and selecting the candidate that maximized the consensus set (inliers), making it particularly effective against gross contamination \cite{barath2022learning}.
For the study of parametric methods, She and Owen \cite{she2011outlier} originally investigated outlier detection through the lens of penalized regression. Building on this sparsity-based perspective, recent work pushed the boundaries of high-dimensional robust regression. For instance, Shen et al. \cite{shen2025computationally} proposed algorithms achieving statistical optimality under heavy-tailed noise. In the field of machine learning, Wan et al. \cite{wanj2024enhanced} utilized an adaptive robust loss function (ARLF) to replace traditional loss functions, significantly enhancing model stability and accuracy in complex and noisy data environments.

\subsection{Concept drift understanding and detection methods}
\label{Concept Drift Understanding and Detection Methods}
Concept drift refers to the phenomenon where the joint probability distribution $P(X, y)$ of the data stream changes over time, leading to the degradation of the predictive model \cite{lima2022learning}. 
Generally, drifts are categorized into virtual drift (changes in the feature distribution $P(X)$ without affecting the decision boundary) and real drift (changes in the posterior $P(y|X)$) \cite{agrahari2022concept}. To address these non-stationarities, detection methodologies have evolved from statistical process control to sophisticated distribution monitoring and integrated adaptive frameworks \cite{sharief2024multi} \cite{lukats2024benchmark}. 
While drift detection is well-established in classification domains using error-rate-based algorithms such as Drift Detection Method (DDM) \cite{gama2004learning} \cite{costa2018drift} and Early Drift Detection Method (EDDM) \cite{baena2006early}, these approaches are ill-suited for regression tasks, as they necessitate transforming continuous errors into binary labels via artificial thresholding. This discretization process discards the magnitude of the error—a critical signal in regression drifts—and introduces bias dependent on the threshold definition. Therefore, research in regression drift detection pivoted towards methods that directly analyze the statistical properties of continuous data streams. The Page-Hinkley (PH) test \cite{mouss2004test} \cite{zhan2023unsupervised} is a classic method that monitors the cumulative error sum to detect abrupt shifts; however, it relies heavily on manually tuned thresholds. To automate this, ADWIN \cite{bifet2007learning} employed adaptive windowing based on Hoeffding bounds to detect changes in the mean without user parameters. Addressing the limitation of mean-based detection, KSWIN \cite{raab2020reactive} \cite{hinder2023one} utilized the Kolmogorov-Smirnov test to identify changes in the shape of the data distribution, capturing drifts that do not affect the mean. A method named DataStream Adapt (DSA) \cite{yashwanth2023datastreamadapt} was proposed that couples hybrid error-monitoring with dynamic ensemble reweighting, effectively detect both abrupt and slow drifts.

\section{Methodology}
\label{section:Methodology}

\subsection{Problem setup}\label{Problem setup}
Consider a data stream $\{Z_1, Z_2, ..., Z_{t-1}\}$ within a time interval $[1,t-1]$, where each data instance $Z_i=(X_i,y_i)$ comprises a feature vector $X_i$ and a corresponding label $y_i$. It adheres to a distribution denoted as $P_{t-1}(X, y)$. Concept drift refers to the phenomenon where the data distribution changes over time within a data stream. Upon advancing to $t$ and subsequent timestamps, $P_{t-1}(X, y) \neq P_{t}(X, y)$ signifies the emergence of concept drift. 

Given that the new data $Z_t$ has just arrived, we are unable to definitively classify its state. Consequently, in the process of real-time data stream monitoring, the identification of point outliers or concept drifts inevitably involves a certain degree of temporal lag. This lag stems from the inherent uncertainty of future data, necessitating that we wait for at least one additional time point of data updates before making a relatively accurate determination in the actual evolution of the data stream. 

The core focus problem lies in precisely identifying and distinguishing between point outliers and concept drifts when they may coexist within data streams. In other words, our objective is to accurately determine whether the data at time $t-1$ belongs to the category of a warning point outlier, a confirmed point outlier, an abrupt concept drift, an incremental concept drift, or a normal data point when the new data $Z_t$ arrives.



\subsection{Robust model construction based on dynamic windows}\label{Robust regression learner}
The general form of a linear regression model is represented as $\bm{y} = \bm{X\beta+\epsilon}$, where $\bm{y}\in \mathbb{R}^{w}$ is a vector comprising the response variables, $\bm{X}\in \mathbb{R}^{w\times d}$ is the design matrix formed by features, $\bm{\beta}\in \mathbb{R}^{d}$ is the parameter vector to be estimated, $\bm{\epsilon}\in \mathbb{R}^{w}$ 
 stands for the error vector, and $w$ denotes the size of the reference window. 
 
The Ordinary Least Squares method is commonly used for parameter estimation; however, its high sensitivity to outliers makes it unreliable when dealing with noisy data streams. When confronting the simultaneous challenges of outliers and data drift, the crucial factors for maintaining model performance are the dynamic window update strategy and the selection of the regression model and drift detector.
 
To address these challenges, we recommend using robust regression approaches over the dynamic window. This practice yields more resilient model coefficient estimates and facilitates superior predictive performance. Furthermore, the system presented in this paper minimizes unnecessary overhead by requiring updates only at initialization and upon system-reported drift.

\subsection{Dual-channel decision process}\label{Dual-channel decision process}

The dual-channel decision process involves designing two specific algorithmic blocks, corresponding to two independent channels, to determine whether the new data represents a point outlier or a concept drift by analyzing the trend of prediction residuals.
The reason for proposing the dual-channel decision process is the necessity to simultaneously consider both possibilities during the decision-making process. 

\begin{figure}
\centering
\scalebox{0.28}{
\includegraphics{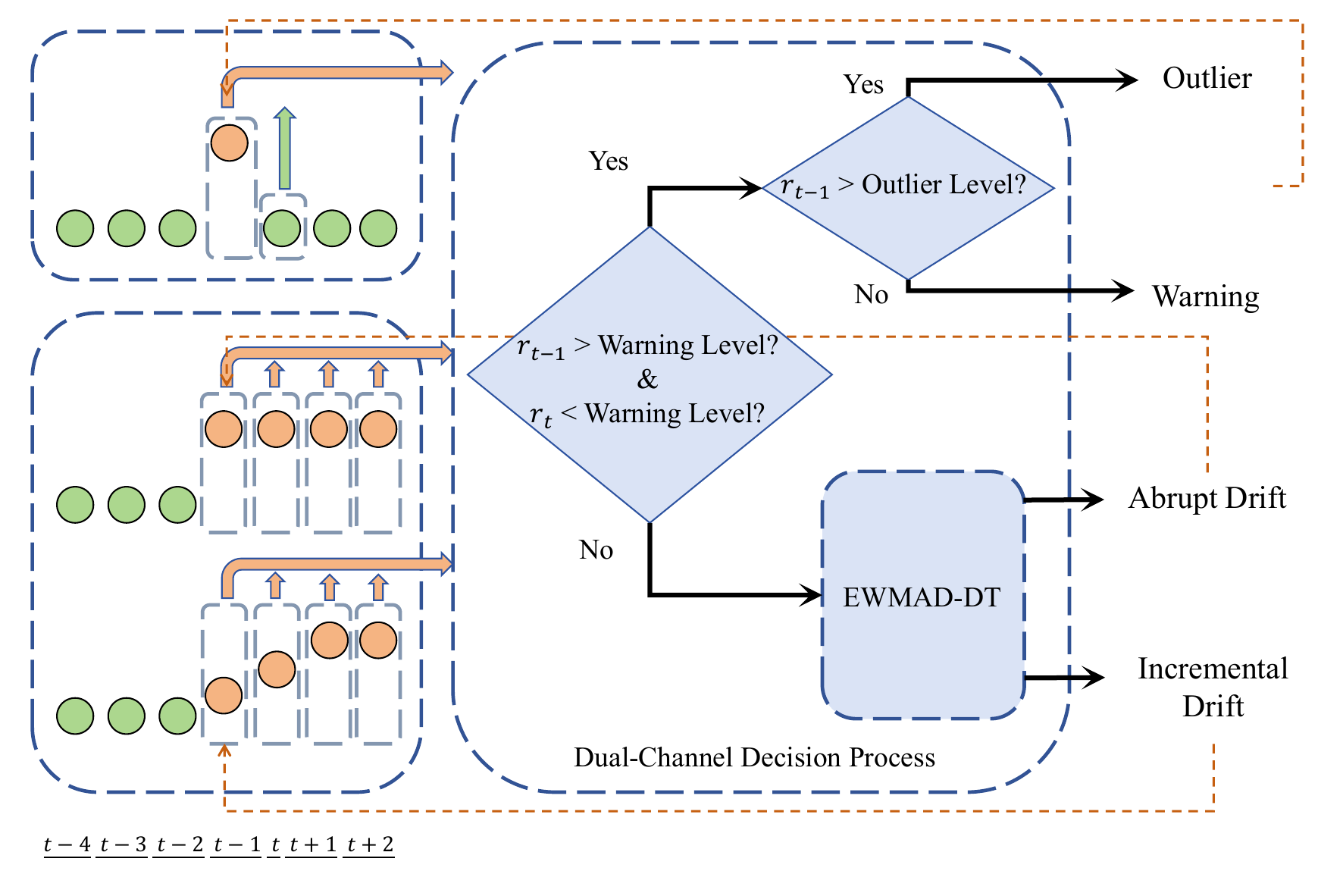}
}
\caption{Simplified diagram of dual-channel decision process. }
\label{fig3:dual}
\end{figure}

We illustrate Fig. \ref{fig3:dual} to enhance understanding. Determining the data type at time $t-1$ relies on analyzing subsequent observations. Point outlier detection follows a simple logic: whenever the residual at $t-1$ surpasses the warning outlier level, the system checks if the residual at time $t$ returns to its normal range.  A return below the warning level confirms an outlier at $t-1$. Failing that, the residual at $t-1$ is channeled for subsequent analysis by the EWMAD-DT drift detector, which is tasked with both detecting drift and categorizing its specific type.

The advantage of adopting this process lies in its ability to accurately distinguish between point outliers and concept drift. When faced with new data exhibiting large prediction residuals, the system can automatically adjust dynamic thresholds to promptly issue warnings for point outliers. Simultaneously, it is capable of detecting persistently poor prediction residuals, thereby swiftly reporting instances of concept drifts. 

\begin{algorithm}[ht]
    \caption{Joint Outlier and Concept Drift Detection}
    \label{alg:Half-KFN statistic}
    \renewcommand{\algorithmicrequire}{\textbf{Input:}}
    \renewcommand{\algorithmicensure}{\textbf{Output:}}
    \begin{algorithmic}[1]
         \scriptsize{\REQUIRE Online multivariate data $\left\{\bm{Z}_{t}\right\}_{t=1}^{\infty}$ = $\left\{(\bm{X}_{t},y_{t})\right\}_{t=1}^{\infty}$ where $\bm{X}_{t} \in \mathbb{R}^{d}$ and $y_{t} \in \mathbb{R}$; an IPOD learner $f: \mathbb{R}^{d} \rightarrow \mathbb{R}$; window size $w$; concept drift alarm parameter $\xi$, $\tau$; difference order $k$; smoothing constant $\tau$  
        \ENSURE \textit{$decision_{t-1}$(normal, warning, outlier or  drift?)}
        \STATE Initialize $r_0 = 0$, set two boolean values $warning_0 = outlier_0 =  False$, $t=1$
        \STATE A new data $Z_t$ arrives in data streams
        \IF{$t=1$ or the state at time $t-1$ is a drift  } 
        \STATE Clear $\overline{r}=0$, $t'=1$, $diff=0$, $m=0$
        \STATE Set regression window $W \leftarrow \{\bm{Z}_{t},...,\bm{Z}_{t+w-1}\}$ 
        \STATE $\bm{X}\leftarrow(\bm{X}_{t},...,\bm{X}_{t+w-1})^T$, $\bm{Y} \leftarrow(y_{t},...,y_{t+w-1})^T$
        \STATE Optimized regression coefficients: $\hat{\bm{\beta}}, \hat{\bm{Y}} \leftarrow regressor$
        \STATE Estimate $\hat{\mu} = mean(\bm{Y} - \hat{\bm{Y}}), \hat{\sigma} = std(\bm{Y} - \hat{\bm{Y}})$
        \ENDIF
        \STATE Calculate the absolute residual $r_t = |y_t - \bm{X_t}\hat{\bm{\beta}}|$
        \STATE $warning_t = |y_t - \bm{X_t}\hat{\bm{\beta}}|>\hat{\mu}+2\hat{\sigma}$
        \STATE $outlier_t = |y_t - \bm{X_t}\hat{\bm{\beta}}|>\hat{\mu}+2.6\hat{\sigma}$
        \IF{$warning_{t-1}$ and not $warning_t$ and the state at time $t-1$ is not a drift}
        \STATE $decision_{t-1} \leftarrow warning$
        \IF{$outlier_{t-1}$}
        \STATE $decision_{t-1} \leftarrow outlier$
        \ENDIF
        \ELSE
        \STATE For data entering the concept drift channel $R_{t'}\leftarrow{r}_{t-1}$
        \STATE Update residual mean $\overline{R}_{t'} \leftarrow \frac{1}{m}\sum_{m=1}^{{t'}}R_{m}$
        \STATE $S \leftarrow (1-\tau)S+\tau(R_{t'}-\overline{R}_{t'})$, $t' \leftarrow t'+1$
        \STATE Update dynamic threshold $\theta \leftarrow \xi \cdot \overline{R}_{t'}$
        \STATE For an upward change $m \leftarrow min(m,S_{t'+1})$
        \STATE $\tilde{S}= S - m$
        \IF{$\tilde{S} \geq \theta$}
        \STATE the state reports a concept drift and reset mechanism
        \STATE $decision_{t-1} \leftarrow drift$ 
        \STATE $\Delta^{k}(\overline{R}_{t'}) \leftarrow \overline{R}_{t'} - \overline{R}_{t'-k}$ if $t'>k$
        \STATE Distinguish drift types based on the sign of $\Delta^{k}(\overline{R}_{t'})$ 
        \ELSE
        \STATE $decision_{t-1} \leftarrow normal$
        \ENDIF
        \ENDIF
        \RETURN $decision_{t-1}$
        }
    \end{algorithmic}
\end{algorithm}

\subsubsection{Point outlier detection}
\label{Point outlier detection}

If a single data instance appears anomalous when compared to the rest of the data, it is designated as a point outlier \cite{singh2012outlier}. Point outliers are the most fundamental and simplest type of outliers and thus have become the focus of much outlier detection research. Based on the assumption of residual normality by the robust regression learner discussed in Section \ref{Robust regression learner}, we estimate the mean and standard deviation of the normal distribution followed by the residuals respectively: 
\begin{equation}
    \hat{\mu} = \frac{\sum\limits_{i=1}^{w}({y_i} - \hat{y}_i)}{w}, \hat{\sigma} = \sqrt{\frac{\sum\limits_{i=1}^{w}({y_i} - \hat{y}_i)^2}{w}}.
\end{equation} 

The absolute value of the prediction residual for new data at $t$ is 
\begin{equation}
    r_t = |y_t - \bm{X_t}\hat{\bm{\beta}}|.
\end{equation}

Thus, the new data can be classified as an outlier by comparing its residual against a threshold derived from a two-sided hypothesis test on the prediction residual. The Prediction Residual Confidence Interval (PRCI), calculated utilizing the sample residuals and a specified confidence level ($\alpha$), serves as the benchmark for identifying point outliers. The PRCI delineates the range within which the residuals of plausible predictions are expected to fall under normal predictive conditions. The prediction level $p=100(1-\alpha)$ determines the critical value from the distribution \cite{hill2010anomaly} \cite{yu2014time}. Assuming that the residuals of the robust regression model adhere to a zero-mean Gaussian distribution, the $p \%$ PRCI can be determined as follows:
\begin{equation}
    \text{PRCI} =\left[ 0,\hat{\mu} + t_{1-\alpha/2,w-1}\times \hat{\sigma} \sqrt{1+\frac{1}{w}}\right]
\end{equation}
where $\hat{\mu}$ and $\hat{\sigma}$ are the mean and standard deviation of the residuals in the sliding window $w$, and $t_{1-\alpha/2,w-1}$ is the critical value of a Student's $t$-distribution with $w-1$ degrees of freedom corresponding to the upper tail probability of $\alpha/2$. The PRCI is referred to as a $t$-interval because it relies on the Student's $t$-distribution, which is suitable for statistical inference when sample sizes are small and the population standard deviation is unknown. When the calculated residual $r_t$ of a new point falls within the PRCI, it is deemed a non outlier; otherwise, it is classified as an outlier.
The advantage of utilizing the PRCI lies in its ability to adapt the interval width based on the prediction level and sample variability.
As the degrees of freedom increase, the $t$-distribution converges to asymptotic normality.

To effectively manage transient anomalies and stochastic fluctuations, we employ a hierarchical filtering mechanism. First, the confirmed outlier filter handles transient, high-magnitude deviations applying immediate exclusion. Second, we define a warning zone for data falling into the buffer area. While the data deviates from the central tendency, this deviation represents stochastic fluctuations or minor interferences. In this state, the system maintains vigilance and filters these samples, yet refrains from triggering an outlier alarm. For a large window size ($w\geq 100$), the warning level ($p=95$) is approximately $\hat{\mu} + 2\hat{\sigma}$, and the confirmed outlier level ($p=99$) is approximately $\hat{\mu} + 2.6\hat{\sigma}$. 

It is important to note that the onset of a gradual drift may initially resemble a confirmed point outlier. In our framework, such ambiguous points may be temporarily flagged as outliers by the rapid channel (Channel I). However, as the drift persists, the cumulative evidence in the EWMAD-DT detector (Channel II) will inevitably trigger a drift alarm, correcting the initial classification. This design prioritizes robustness against transient noise while ensuring that genuine structural shifts are eventually captured.

\begin{figure*}
\centering
\scalebox{0.51}{
\includegraphics{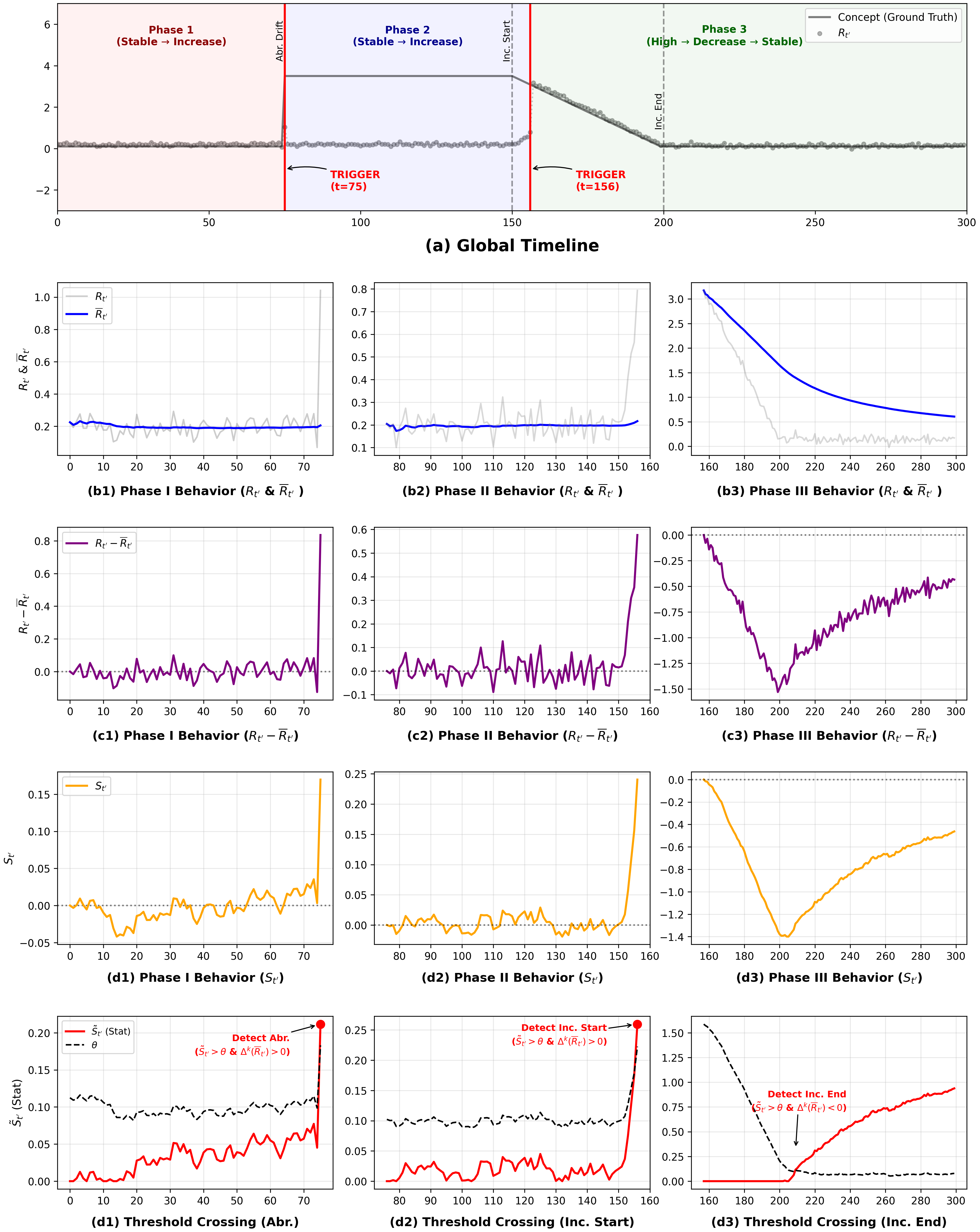}
}
\caption{Schematic illustration of the EWMAD-DT detector and adaptation mechanism. (a) Global view of the ground truth concept and observed data streams during incremental drift. (a1)–(d3) Detailed evolution of statistical metrics ($R_{t'}$, $R_{t'}-\overline{R}_{t'}$, $S_{t'}$, $\tilde{S}_{t'}$) during the monitoring phase I, II and III.}
\label{fig:EWMAD-DT)}
\end{figure*}

\subsubsection{Concept drift detection}
\label{Concept drift detection}

Only when the prediction residual of new data fails to meet the criteria for the point outlier detection channel will it proceed to the concept drift detection channel. To monitor for concept drift, we proposed a new method named Exponentially Weighted Moving Absolute Deviation with Distinguishable Types (EWMAD-DT). The visualization process of our proposed detector is illustrated in the Fig. \ref{fig:EWMAD-DT)}. This method tracks the growth of prediction residuals to signal deviations. We introduce a weighted cumulative variable, which is defined as follows:
\begin{equation}
    S_{t'+1} = (1-\tau)S_{t'}+\tau\left(R_{t'}-\overline{R}_{t'}\right)
\end{equation}
where $R_{t'}$ is the absolute value of the prediction residual for data entering the concept drift channel at the $t-1$ step, $\overline{R}_{t'} = \frac{1}{m}\sum_{m=1}^{{t'}}R_{m}$ is the average of the absolute values of historical residuals since reset, and $\tau$ corresponds to the smoothing constant. 
The core of this approach lies in the fact that concept drift within the data stream results in substantial increases in $R_{t'}-\overline{R}_{t'}$. 

Both abrupt drift and the starting point of incremental drift can lead to an increase in $R_{t'}-\overline{R}_{t'}$. Naturally, a well-defined threshold is sufficient to signal the onset of drift when the monitoring statistic $S_{t'}$ exceeds it.  Many existing methods focus solely on detection rather than differentiation. However, our subsequent improvement precisely addresses this gap, enabling us to distinguish between abrupt and incremental drifts. 
We observe that when an incremental drift occurs, it triggers the reset mechanism to re-update the model. The new training window (with a size at least twice that of the incremental drift process) encompasses the incremental drift phase and the subsequent stable data stream.
Within this window, $R_{t'}$ first decreases and then remains constant. This precisely results in an increase in $R_{t'}-\overline{R}_{t'}$ (recovering from negative values towards zero) after the incremental drift ends. Therefore, $S_{t'}$ can increase in three scenarios: at the abrupt drift, at the start of incremental drift, and after incremental drift ends. To minimize detection delay, we refine $S_{t'}$ by subtracting its running minimum, enabling us to precisely identify the exact end of an incremental drift. This leads to our formally defined statistic:
\begin{equation}
    \tilde{S}_{t'} = S_{t'} - \min\limits_{1 \leq m \leq t'}S_{m}.
\end{equation}

We also involve setting a dynamically adjusted threshold. A concept drift is reported when the above value exceeds a dynamic threshold $\Theta_{t'} = \xi \cdot \theta_{t'}$, where $\xi \in (0,1)$ controls sensitivity, where $\xi$ is a specific constant value that controls the sensitivity to drift by adjusting the threshold, and $\theta_{t'} = (1-\tau)\theta_{t'-1} + \tau\overline{R}_{t'-1}$. In data stream processing, when a drift is detected, variables are reset to focus more on recent data, thereby reducing response time and lag effects. 

The above process can effectively detect the three scenarios. To further distinguish between them, we observe that at the abrupt drift, at the start of incremental drift, $\overline{R}_{t'}$ exhibits an increasing trend, whereas at the end of incremental drift, the variation in $\overline{R}_{t'}$ first decreases and then stabilizes. To capture this, we define the $k$-th order difference as follows:
\begin{equation}
    \Delta^{k}(\overline{R}_{t'}) = \overline{R}_{t'} - \overline{R}_{t'-k}, \quad t'>k.
\end{equation}
By examining the sign of $\Delta^{k}(\overline{R}_{t'})$, we can determine whether $\overline{R}_{t'}$ is increasing or decreasing. Specifically, the last detected moment where $\Delta^{k}(\overline{R}_{t'})>0$ also occurs before the end of incremental drift is identified as the start of incremental drift, while all other detected drifts are classified as abrupt, thereby enabling the distinction between abrupt and incremental drifts.

The method offers significant advantages: straightforward implementation, low computational complexity for large-scale data streams, the capability to differentiate between abrupt and incremental drift, and fast sample evaluation that enhances both real-time performance and monitoring efficiency.

\section{Experimental evaluation}\label{section:Experimental evaluation}
In this section, we use different robust regression learners, and illustrate the advantages of our proposed framework by comparing EWMAD-DT with other drift detection methods for data streams. The evaluations involved different scenarios using both synthetic and real-world datasets. The synthetic dataset enables us to control relevant parameters and empirically evaluate algorithms using streaming data of different types, including outliers, abrupt drifts, and incremental drifts. The synthetic data used in this section comprise artificial datasets with anomalies and various types of drifts, as well as real-world datasets with injected anomalies and different types of drifts. In addition, real-world datasets are also employed in this section, allowing us to assess the advantages of our proposed method in practical scenarios.
The source code is available online\footnote{\href{https://github.com/bbwang1030/EWMAD-DT}{https://github.com/bbwang1030/EWMAD-DT}}.

\subsection{Related robust regression learner}
\textbf{$\Theta$-IPOD} considers the model $\bm{y} = \bm{X\beta+\gamma+\epsilon}, \bm{\epsilon} \sim \mathcal{N}(0,\sigma^2\bm{I})$
 imposing sparsity constraints on $\bm{\gamma}\in \mathbb{R}^{w}$, in which the parameter $\gamma_i \neq 0$ indicates that the $i$-th observation is an outlier. The objective is to consider optimizing the following:
 \begin{equation}
 \label{opt}
     f_P(\bm{\beta},\bm{\gamma})\equiv \frac{1}{2}\|\bm{y}-\bm{X}\bm{\beta}-\bm{\gamma} \|_2^2  + \sum\limits_{i=1}^{w}P(\gamma_i;\lambda_i),
 \end{equation}
where $\lambda_i$ are a collection of penalty parameters. A threshold function $\Theta(\cdot; \lambda)$ is coupled with $P(\theta;\lambda) = \int_0^{\left|\theta\right|}(sup\{t:\Theta(t;\lambda)\leq u\}-u) du+q(\theta;\lambda)$, where $q(\cdot;\lambda)$ is nonnegative and $q(\Theta(\theta;\lambda))=0$ for all $\theta$. She \cite{she2011outlier} obtained robust estimates of $\bm{\beta}$ and $\bm{\gamma}$ through an iterative algorithm.  

\textbf{RANSACRegressor (RANSAC)} begins by randomly selecting the minimum number of samples required to fit an initial linear model. This model is then used to test all other data points in the dataset. Based on a predefined residual threshold, the data are classified into ``inliers'' and ``outliers''. The model is subsequently refitted using only the inliers. Through iterative random sampling and refitting, the final model is selected as the one with the largest number of inliers. This approach is highly robust against outliers and can effectively handle data heavily contaminated with a high proportion of outliers.

\textbf{HuberRegressor (Huber)} employs the Huber loss function as its optimization objective. For data points with small residuals, it applies the squared loss, while for points with large residuals, it switches to the linear loss term $\delta|y_t-\hat{y}_t|-\delta^2/2$. This mechanism prevents outliers from exerting excessive influence, as would occur with the squared loss, thereby mitigating the impact of anomalies. This approach combines the efficiency of OLS with the robustness of absolute error loss, representing a balanced trade-off between the two.

\textbf{TheilSenRegressor (TheilSen)} is a non-parametric linear regression model based on the Theil-Sen estimator. This estimator determines the final regression slope by calculating the median of the slopes between all pairs of data points. Subsequently, the intercept is determined using the median of the intercept residuals. This method is known to perform stably when the data contain outliers.

\textbf{ARLF} constructs a generalized framework that unifies and smoothly transitions between distinct loss mechanisms By introducing a tunable shape parameter $\alpha$. Specifically, the function degrades to the L$2$ loss (MSE) when $\alpha$ approaches $2$, behaves as the Charbonnier loss when $\alpha$ is set to $1$, and suppresses the influence of outliers by down-weighting large errors when $\alpha\le0$. When $\alpha$ decreases to $-2$ or approaches negative infinity, the function behaves as the Geman-McClure loss or the Welsch loss. Crucially, this method features adaptivity by treating $\alpha$ as a learnable parameter that is automatically optimized during training. This mechanism enables the model to autonomously determine the requisite level of robustness based on the underlying data distribution, significantly enhancing its resilience in complex data environments.

The aforementioned algorithm provides a robust regression learner, paving the way for our subsequent work in simultaneously identifying point outliers and concept drifts.

\subsection{Related drift detector}
\textbf{ADaptive Windowing (ADWIN)} operates by maintaining an adaptively-sized sliding window for drift detection \cite{bifet2007learning}. The algorithm continuously tests whether a split point exists that can divide the current window into two subsets such that a statistical measure shows a statistically significant difference between them. Once such a split point is detected, the algorithm confirms the occurrence of concept drift and promptly discards all historical data prior to that point from the window.



\textbf{Kolmogorov-Smirnov Windowing (KSWIN)} maintains a sliding window of fixed size $w_0$. The last $s_0$ samples of the window are assumed to represent the last concept considered as $R$. From the first $w_0-r_0$ samples, $r_0$ samples are uniformly drawn, representing an approximated final concept $W$. The KS-test is performed on the windows $R$ and $W$ of the same size. The KS-test compares the distance of the empirical cumulative data distribution $d(R,W)$, and a concept drift is detected if $d(R,W)>\sqrt{ln(\alpha)/r_0}$ \cite{raab2020reactive}.

\textbf{Page-Hinkley (PH)} considers introducing a cumulative variable $\text{CumDiff}_{t'} = \sum_{m=1}^{t'} (r_{t'}-\overline{r}_{t'} -\tau)$, where $\tau$ corresponds to the magnitude of the permissible difference. A PH statistic defined as $\text{PH} = \text{CumDiff}_{t'} - \min_{1\leq m\leq t'}\text{CumDiff}_{m}$. A concept drift is reported when the above value exceeds a threshold.

\textbf{DataStream-Adapt (DSA)} is a unified and adaptive framework designed to detect and respond to both abrupt and slow concept drifts in real-time data streams \cite{yashwanth2023datastreamadapt}. The method is designed to leverage two mechanisms: the Cumulative Sum (CUSUM) for identifying abrupt shifts, whereas lagged moving averages are utilized to capture slow drifts. Two corresponding thresholds need to be set for the detection of abrupt drift and slow drift.

\subsection{Datasets}

\textbf{Abrupt Drift Artificial Dataset (Abr$_{\delta}$)} consists of $50000$ 10-dimensional data samples denoted as $\{X_t,y_t\}_{t=1}^{50000}$. Data are generated from the model:
$y_t = \bm{\beta}_t \bm{X}_t +\gamma_t+\epsilon_t$, where the predictor matrix $\bm{X}_t \sim \mathcal{U}_{10}{[0.2,0.5]}$ and noise ${\epsilon_t} \sim \mathcal{N}(0,0.001)$. The stream is divided into 50 segments, each containing $1000$ samples. 
For the $i$-th segment ($i=1,\dots,50$), a target coefficient vector $\bm{\beta}^{(i)}$ is sampled from $\mathcal{U}_{10}[0, 1]$. In the abrupt setting, the coefficient vector updates instantaneously at the boundary: $\bm{\beta}_t = \bm{\beta}^{(i)}$ for $t \in (1000(i-1), 1000i]$.
Point outliers are injected with probability $\delta$. If $t$ is an outlier index, an offset $\gamma_t$ is added, drawn from $\mathcal{N}(\pm a, b)$ where $a \sim \mathcal{U}[0.5, 1]$, $b \sim \mathcal{U}[0, 0.1]$, and the sign is equiprobable; otherwise, $\gamma_t = 0$. 

\textbf{Incremental Drift Artificial Dataset (Inc$_{\delta}$)} simulates a smooth transition between concepts. Unlike the instantaneous change in Abr$_{\delta}$, the coefficients linearly interpolate from the previous state $\bm{\beta}^{(i-1)}$ to the new target $\bm{\beta}^{(i)}$ over a transition window of length $L$ (e.g., $L=50$). Specifically, for the $i$-th segment starting at $T_{i} = 1000(i-1)$, the coefficient $\bm{\beta}_t$ is defined as:
\begin{equation}
\bm{\beta}_t = \begin{cases}
\bm{\beta}^{(i-1)} + \frac{t-T_{i}}{L}(\bm{\beta}^{(i)} - \bm{\beta}^{(i-1)}), & \text{if } T_{i} < t \leq T_{i}+L  \\
\bm{\beta}^{(i)}, & \text{if } T_{i} + L<t
\end{cases}
\end{equation}
This formulation ensures a gradual evolution of the concept before stabilizing. Other settings remain consistent with Abr$_{\delta}$. 

\textbf{Mixed Drift Artificial Dataset (Mix$_{\delta}$)} incorporates both drift patterns. For each segment $i$, a Bernoulli trial determines the drift type: with probability $0.5$, the transition is abrupt (as in Abr$_{\delta}$), and with probability $0.5$, it is incremental (as in Inc$_{\delta}$).

\textbf{Bike Sharing Real Dataset\footnote{\href{https://archive.ics.uci.edu/dataset/275}{https://archive.ics.uci.edu/dataset/275}} (Bike)} comprises 732 real-world daily records of bicycle rentals over a two-year period, featuring four primary explanatory variables including temperature, feeling temperature, humidity, and wind speed, as well as a target variable that represents the total rental bike count. This dataset is non-stationary and inherently contains outliers.

\textbf{GNFUV Unmanned Surface Vehicles Sensor Real Dataset\footnote{\href{https://archive.ics.uci.edu/dataset/452}{https://archive.ics.uci.edu/dataset/452}} (pi2,pi3,pi4,pi5)} records sensor readings from four Unmanned Surface Vehicles (USVs), utilizing humidity as the input feature and temperature as the regression target. Due to the dynamic environmental conditions of the testbed, the dataset exhibits natural concept drift and outliers, making it an ideal candidate for evaluating outlier and drift detection in regression tasks. The four subsets contain $1532$, $899$, $1766$, and $2078$ samples, respectively.

\textbf{Worker Productivity Real Dataset\footnote{\href{https://archive.ics.uci.edu/dataset/597}{https://archive.ics.uci.edu/dataset/597}} (Worker)}
comprises 691 valid instances after removing missing values, and six main feature variables are selected: targeted productivity, Standard Minute Value, Work in Progress, amount of overtime, amount of financial incentive, and number of workers, as well as the target variable, actual productivity.


\textbf{Air Quality\footnote{\href{https://archive.ics.uci.edu/dataset/360}{https://archive.ics.uci.edu/dataset/360}} (Airquality)} comprises 9358 instances of hourly averaged responses from an array of 5 metal oxide chemical sensors embedded in an Air Quality Chemical Multisensor Device. The data was collected from March $2004$ to February $2005$ at the road level in a significantly polluted area. After discarding instances with missing values, $827$ valid samples remained. The model utilizes $13$ distinct features to forecast the NO2 concentration levels.

\subsection{Evaluation metric}
In the simulation of artificial datasets for the identification of point outliers or concept drifts, we categorize the experimental results shown in Table \ref{table:1}. 

\begin{table}[ht]   
\begin{center}   
\caption{A $2\times2$ confusion matrix for outlier and concept drift detection.
}  
\setlength{\tabcolsep}{1em}  
\scriptsize
\label{table:1} 
\begin{threeparttable}
\centering
\begin{tabular}{c c c c} 
\toprule
  & & \multicolumn{2}{c}{Detection}  \\ 
 & & 1 & 0\\
\midrule
\multirow{2}{*}{Truth} & 1 & True positives (TP) & False negatives (FN)\\
& 0 & False positives (FP) & True negatives (TN) \\
\bottomrule
\end{tabular} 
\begin{tablenotes}    
\scriptsize               
\item Note: $1$ for outliers/concept drifts and $0$ for non-outliers/non-drifts.
\end{tablenotes}
\end{threeparttable}
\end{center}
\end{table}

According to these definitions, the positive predicted value ($P_{ppv} = \frac{TP}{TP+FP}$) refers to the proportion of correctly classified positive samples among the total samples classified as positive detections. Recall, also known as sensitivity or true positive rate ($P_{tpr} = \frac{TP}{TP+FN}$), is the proportion of correctly classified positive samples among the actual positive samples. The F1-score is the harmonic mean of $P_{ppv}$ and $P_{tpr}$, used to provide a balanced assessment of a model's performance between precision and recall. Its formula is shown as follows:
\begin{equation}
    \text{F1-score} = \frac{2P_{ppv} \cdot P_{tpr}}{P_{ppv} + P_{tpr}}.
\end{equation}

It is widely acknowledged that real (non-synthetic) data streams with well-documented labeled concept drifts and labeled outliers are rare \cite{demvsar2018detecting}. Although the concepts in these datasets often exhibit instability over an extended period, for the purpose of algorithm evaluation, labeled concept drifts and outliers in synthetic datasets are predominantly introduced artificially. In real-world scenarios, it is usually not feasible to evaluate the accuracy of algorithms for outlier detection and drift detection. However, we can assess the performance of algorithms through the model's Mean Absolute Percentage Error (MAPE):
\begin{equation}
    \text{MAPE} = \frac{1}{n}\sum\limits_{t=1}^{n}\frac{r_t}{y_t},
\end{equation}
where $n$ is the number of instances. It shows the measure of the precision of the estimated values compared to the actual values.
Practically, we emphasize and show the $\text{MAPE}^*$ value calculated after eliminating the detected outliers and concept drifts. This evaluation metric can effectively demonstrate the model's fitting performance following the identification of outliers and concept drifts, and is therefore applicable to comparative research on algorithms.

\begin{figure*}[H]  
    \centering    
    \subfigure[Drift/Outlier Detection Performance on Abr$_{\delta=0.01}$] 
    {
        \begin{minipage}[t]{0.85\textwidth}
            \centering          
            \includegraphics[width=1\textwidth]{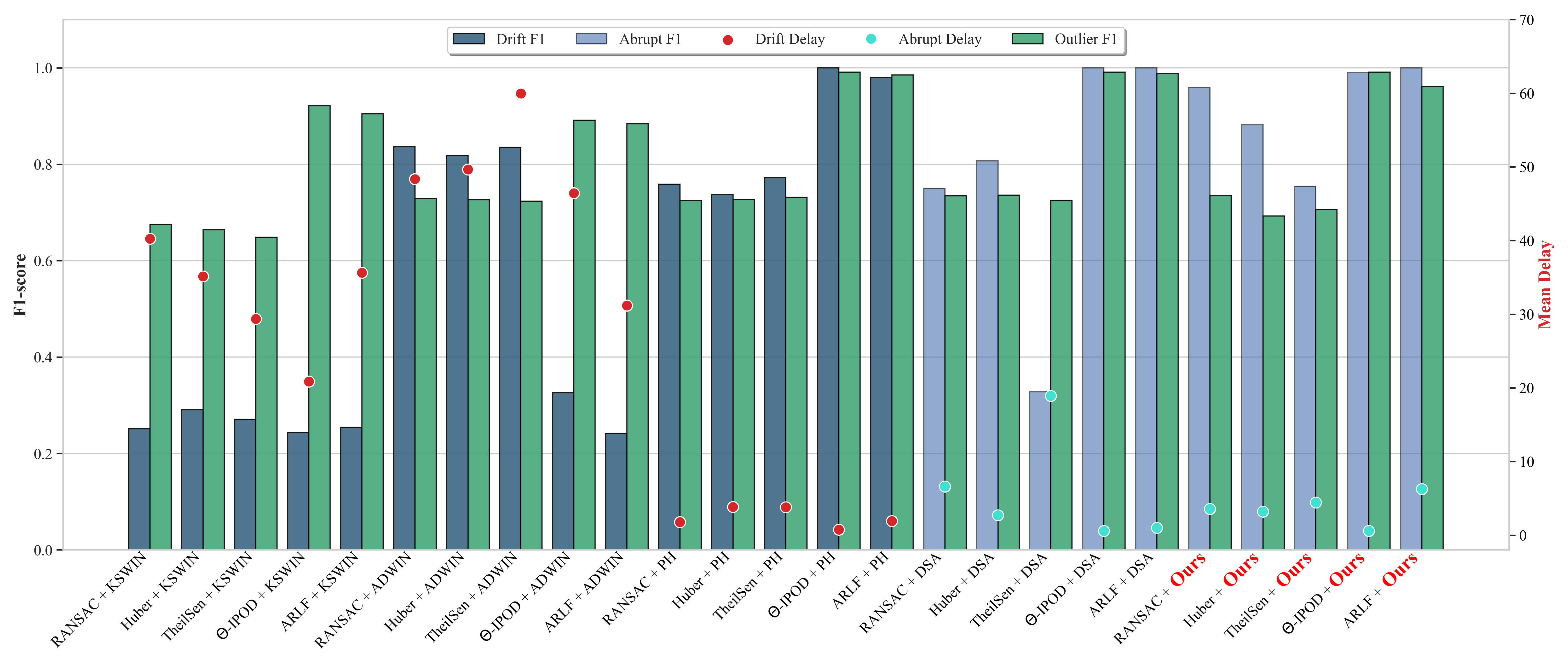}   
        \end{minipage}%
    }
    
    \vspace{-3mm}
    \subfigure[Drift/Outlier Detection Performance on Abr$_{\delta=0.02}$] 
    {
        \begin{minipage}[t]{0.85\textwidth}
            \centering          
            \includegraphics[width=1\textwidth]{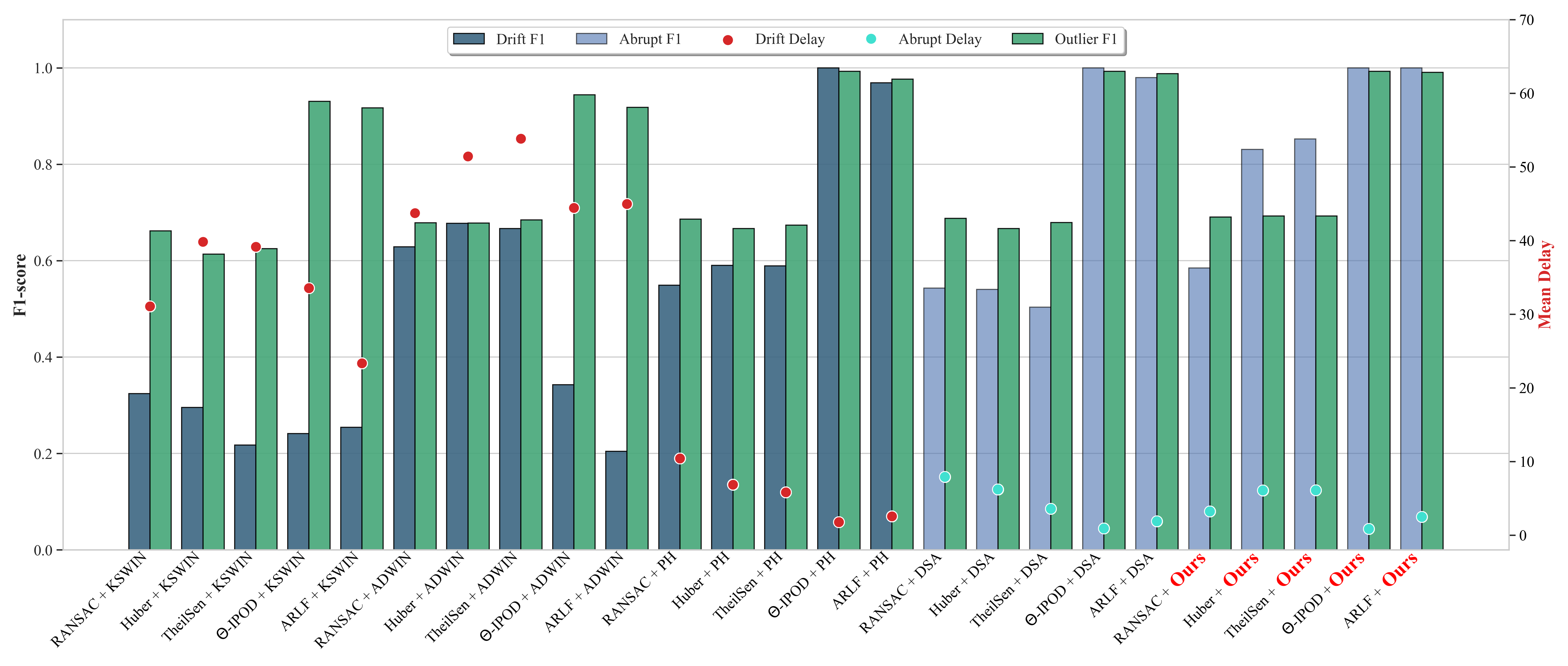}   
        \end{minipage}%
    }
    
    \caption{\small Performance Comparison: Abrupt Drift Detection Accuracy $\&$ Delay on Abr$_{\delta}$ dataset. } 
    \label{fig:abr}  
\end{figure*}

To evaluate the timeliness of the proposed method, we employ the mean delay of drift detection as a key performance metric. Mean delay quantifies the average number of instances processed by the system from the moment a concept drift occurs until it is successfully identified. Let $\mathcal{T} = \{t_1, t_2, \dots, t_m\}$ denote the set of ground truth drift points in a synthetic data stream. For each ground truth drift $t_i$, we search for the first subsequent detection alert $\hat{t}_i$ generated by the algorithm within a predefined tolerance window $c$ (i.e., $t_i \le \hat{t}_i \le t_i + c$). The detection delay for the $i$-th drift is defined as $\delta_i = \hat{t}_i - t_i$. The mean delay is then calculated over all correctly detected drifts:
\begin{equation}
    \overline{\text{Delay}} = \frac{1}{TP} \sum_{j=1}^{TP} \text{Delay}_j,
\end{equation}
where $N_{TP}$ is the total number of true positive detections. A lower mean delay indicates a faster response to concept changes, which is critical for minimizing the accumulation of prediction errors during the transition phase. In our experiments, we set the tolerance $c=100$.

\subsection{Parameter tuning}
During the process of parameter optimization for the algorithm, we opt for the parameter combination that yields the minimum MAPE. In various drift detectors, parameter tuning is a necessary step. Sequential Uniform Design (SeqUD) \cite{yang2021hyperparameter}, as a dynamically optimized experimental design method in this paper, integrates the space-filling properties of uniform design with a sequential strategy. It operates through phased iterations and progressively narrows the search region to efficiently locate the optimal solution. Compared to grid search and Bayesian optimization, SeqUD significantly improves optimization efficiency and conserves resources while effectively mitigating the curse of dimensionality. Relative to random search, sequential random strategies, and Sobol sequences, it offers better space-filling properties and more stable results. Methods such as GP-EI, SMAC, and TPE are designed to select new design points one by one, which can lead to inefficient use of computing resources—a drawback SeqUD avoids. Moreover, SeqUD exhibits strong global exploration capability, making it particularly suitable for computationally expensive or high-dimensional complex optimization problems.

\begin{figure*}[H]  
    \centering    
    \subfigure[Drift/Outlier Detection Performance on Inc$_{\delta=0.01}$] 
    {
        \begin{minipage}[t]{0.85\textwidth}
            \centering          
            \includegraphics[width=1\textwidth]{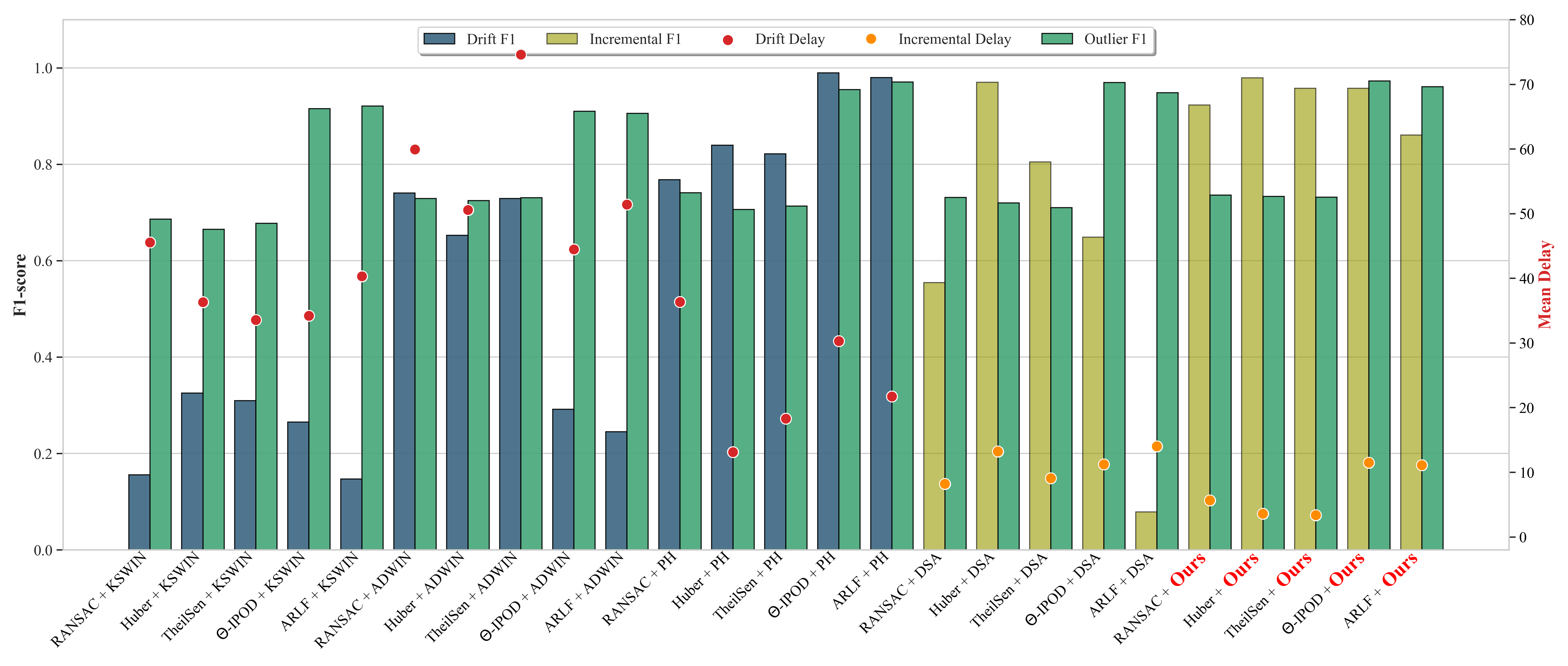}   
        \end{minipage}%
    }
    
    \vspace{-3mm}
    \subfigure[Drift/Outlier Detection Performance on Inc$_{\delta=0.02}$] 
    {
        \begin{minipage}[t]{0.85\textwidth}
            \centering          
            \includegraphics[width=1\textwidth]{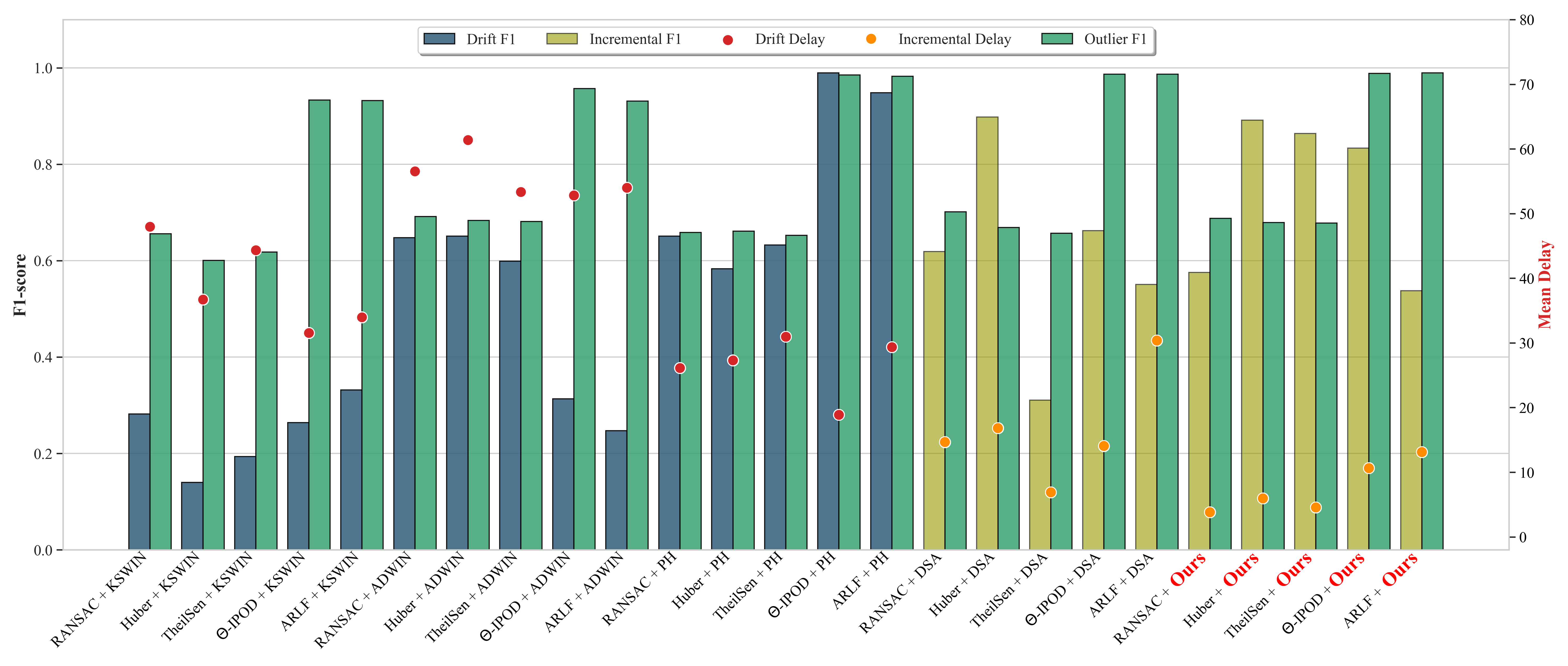}   
        \end{minipage}%
    }
    
    \caption{\small Performance Comparison: Incremental Drift Detection Accuracy $\&$ Delay on Inc$_{\delta}$ dataset. } 
    \label{fig:inc}  
\end{figure*}

\begin{figure*}[ht]  
    \centering    
    \subfigure[Drift/Outlier Detection Performance on Mixed$_{\delta=0.01}$] 
    {
        \begin{minipage}[t]{0.65\textwidth}
            \centering          
            \includegraphics[width=1\textwidth]{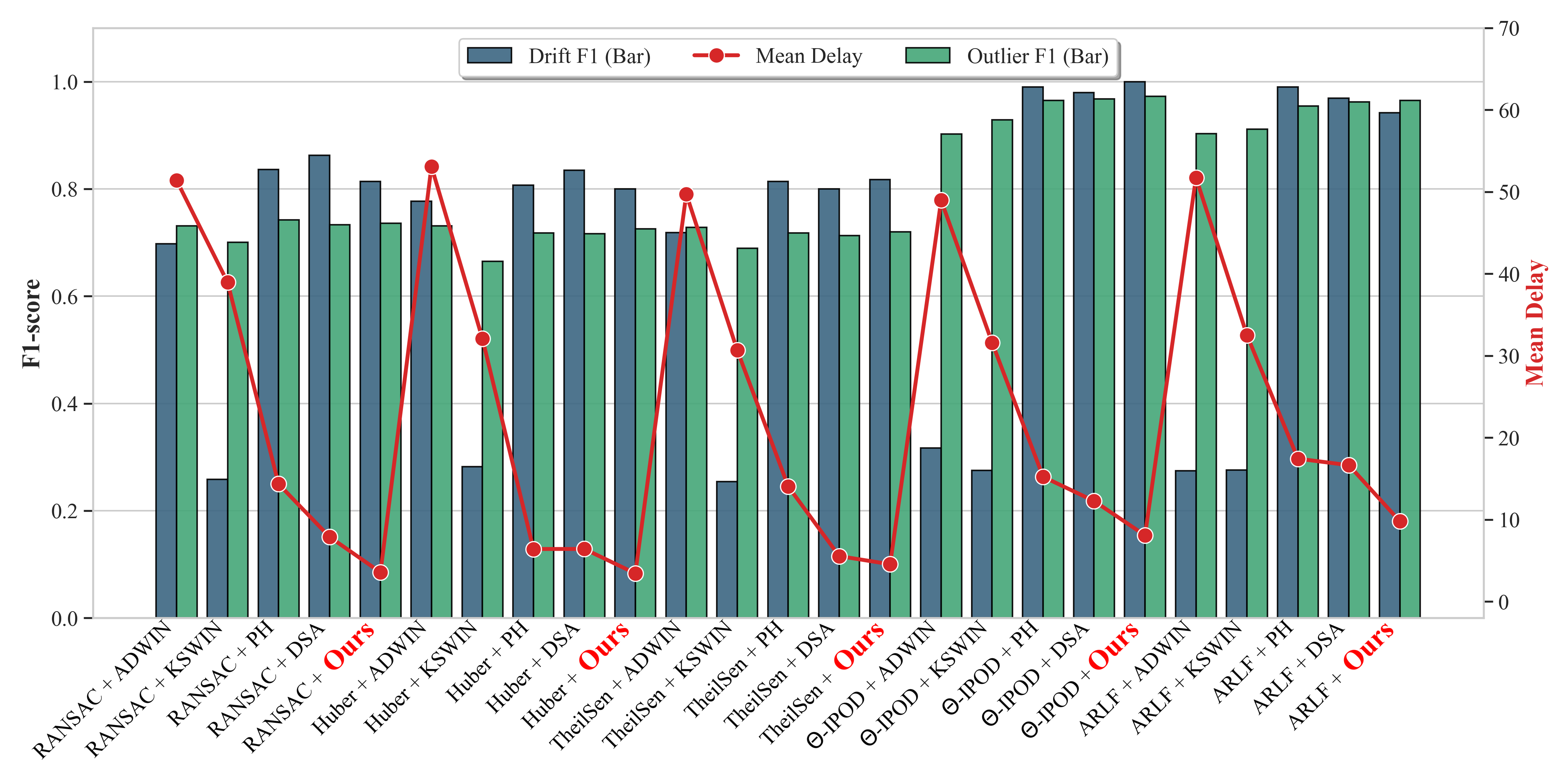}   
        \end{minipage}%
    }
    \subfigure[Detailed Drift Performance on Mixed$_{\delta=0.01}$] 
    {
        \begin{minipage}[t]{0.35\textwidth}
            \centering          
            \includegraphics[width=1\textwidth]{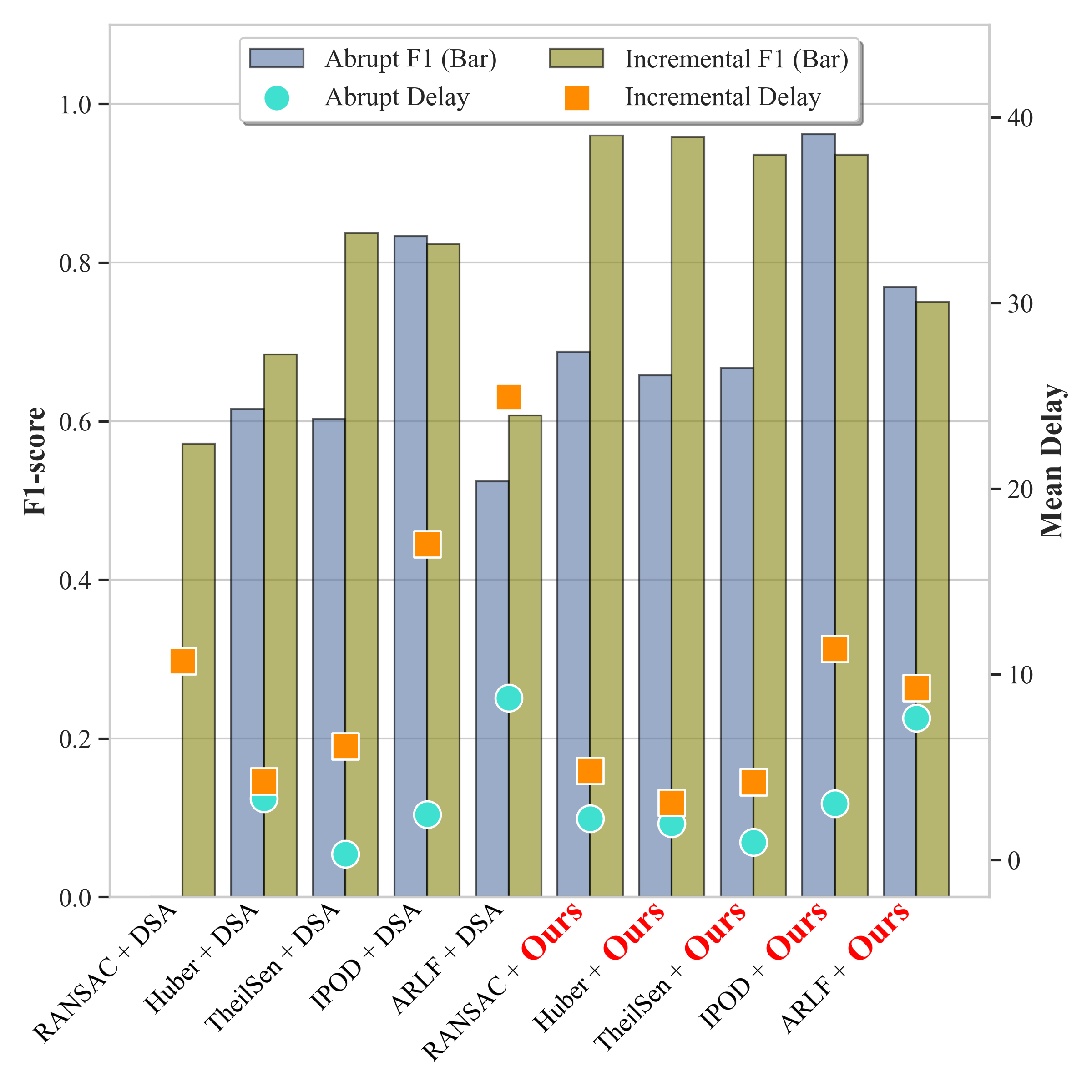}   
        \end{minipage}%
    }

    \vspace{-3mm}
    \subfigure[Drift/Outlier Detection Performance on Mixed$_{\delta=0.02}$] 
    {
        \begin{minipage}[t]{0.65\textwidth}
            \centering          
            \includegraphics[width=1\textwidth]{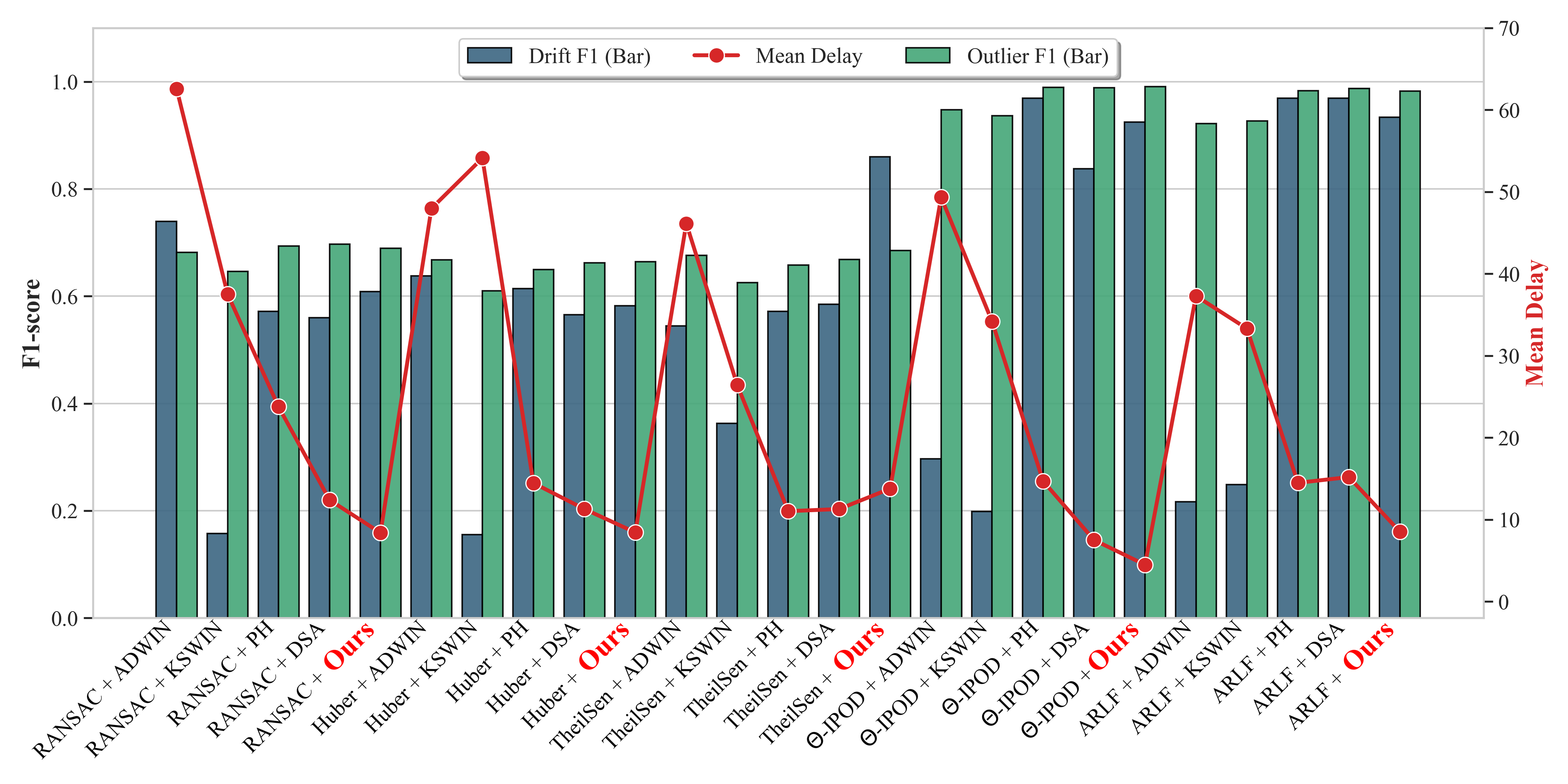}   
        \end{minipage}%
    }
    \subfigure[Detailed Drift Performance on Mixed$_{\delta=0.02}$] 
    {
        \begin{minipage}[t]{0.35\textwidth}
            \centering          
            \includegraphics[width=1\textwidth]{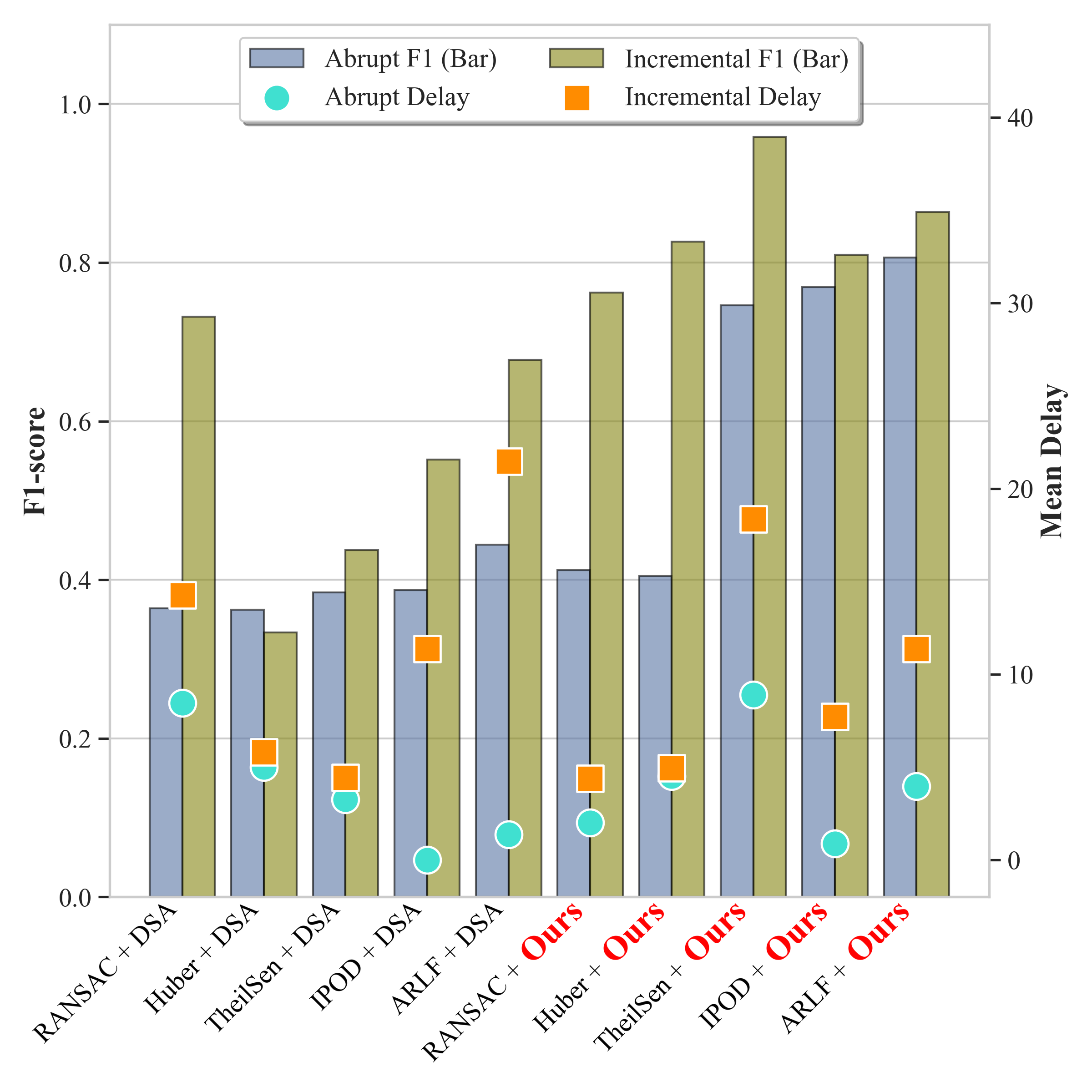}   
        \end{minipage}%
    }
    
    \caption{\small Performance Comparison on Mixed$_{\delta}$ dataset: (a,c) Drift/Outlier Detection Accuracy $\&$ Drift Delay; (b,d) Abrupt/Incremental Drift Detection Accuracy $\&$ Delay. } 
    \label{fig:mix}  
\end{figure*}

\newcommand{\legendline}[2]{%
    \tikz[baseline=-0.5ex]\draw[#1, line width=1.5pt] (0,0) -- (0.8,0);\hspace{3pt}#2%
}

\begin{figure*}[ht]  
    \centering    
    \subfigure[Huber+ADWIN] 
    {
        \begin{minipage}[t]{0.19\textwidth}
            \centering          
            \includegraphics[width=1\textwidth]{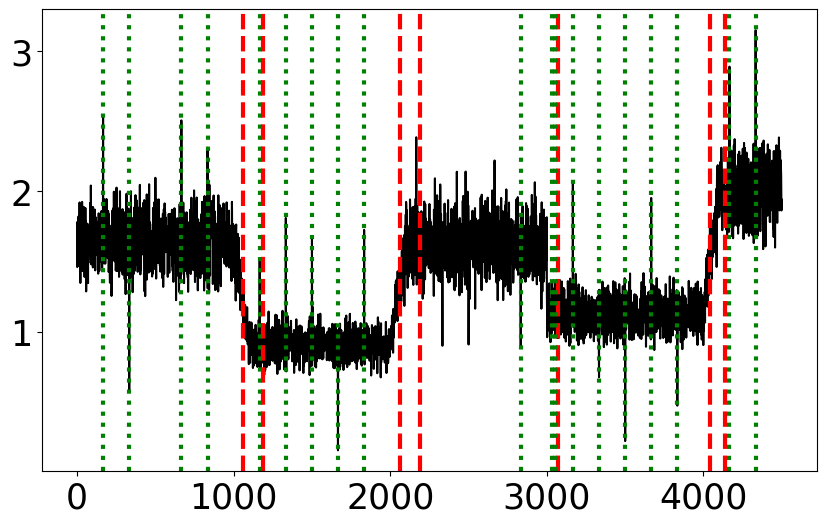}   
        \end{minipage}%
    }
    \subfigure[TheilSen+ADWIN] 
    {
        \begin{minipage}[t]{0.19\textwidth}
            \centering          
            \includegraphics[width=1\textwidth]{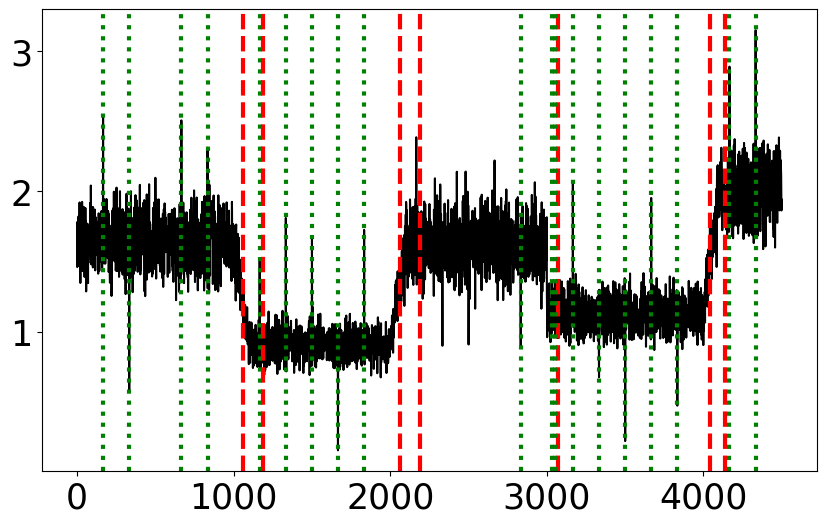}   
        \end{minipage}%
    }
    \subfigure[RANSAC+ADWIN] 
    {
        \begin{minipage}[t]{0.19\textwidth}
            \centering          
            \includegraphics[width=1\textwidth]{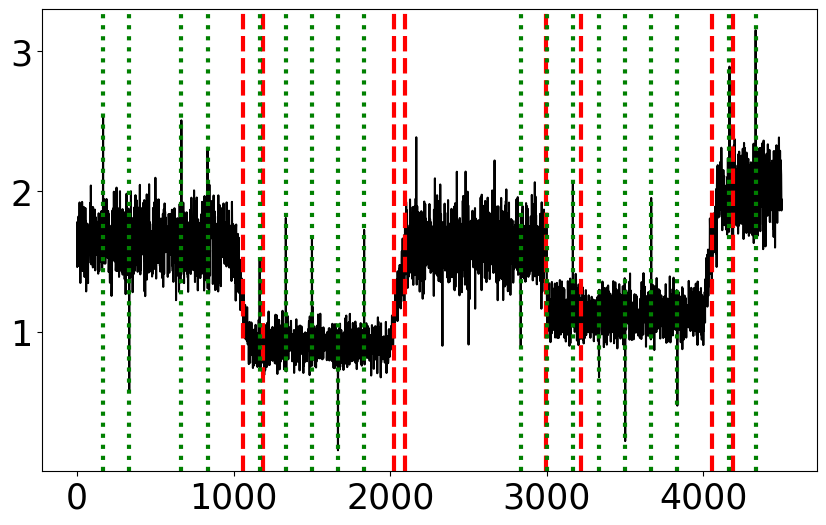}   
        \end{minipage}%
    }
    \subfigure[$\Theta$-IPOD+ADWIN] 
    {
        \begin{minipage}[t]{0.19\textwidth}
            \centering          
            \includegraphics[width=1\textwidth]{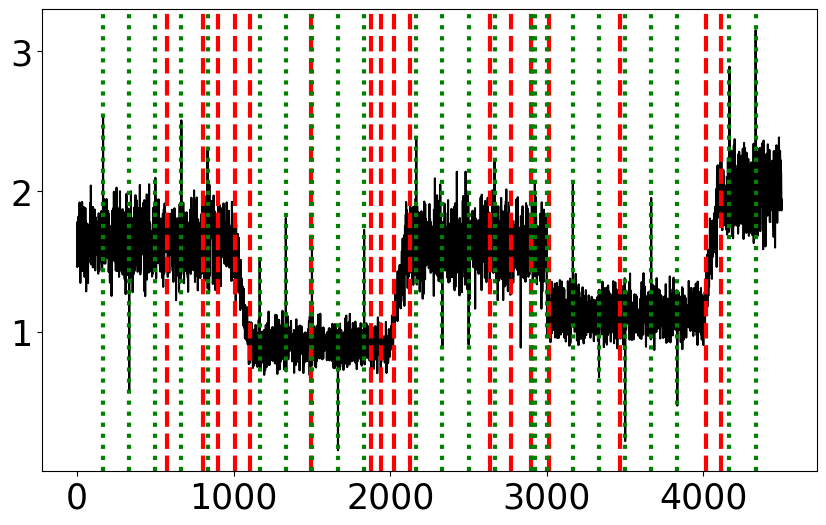}   
        \end{minipage}%
    }
    \subfigure[ARLF+ADWIN] 
    {
        \begin{minipage}[t]{0.19\textwidth}
            \centering          
            \includegraphics[width=1\textwidth]{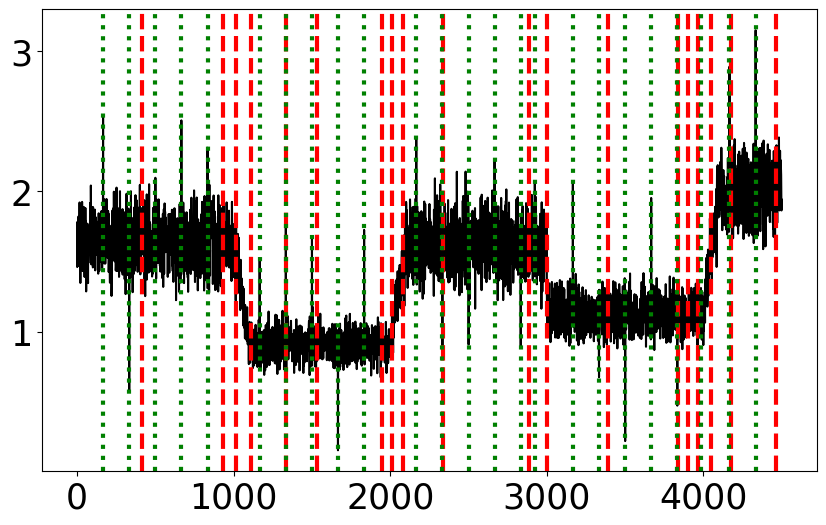}   
        \end{minipage}%
    }

    \subfigure[Huber+KSWIN] 
    {
        \begin{minipage}[t]{0.19\textwidth}
            \centering          
            \includegraphics[width=1\textwidth]{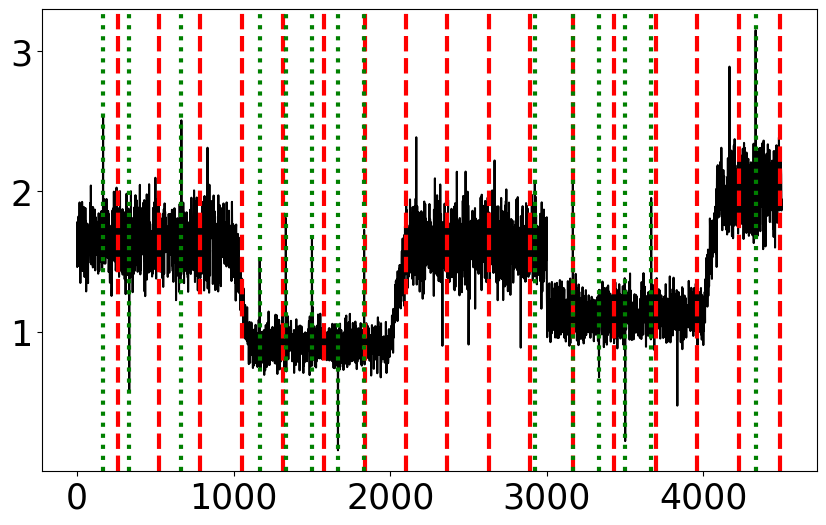}   
        \end{minipage}%
    }
    \subfigure[TheilSen+KSWIN] 
    {
        \begin{minipage}[t]{0.19\textwidth}
            \centering          
            \includegraphics[width=1\textwidth]{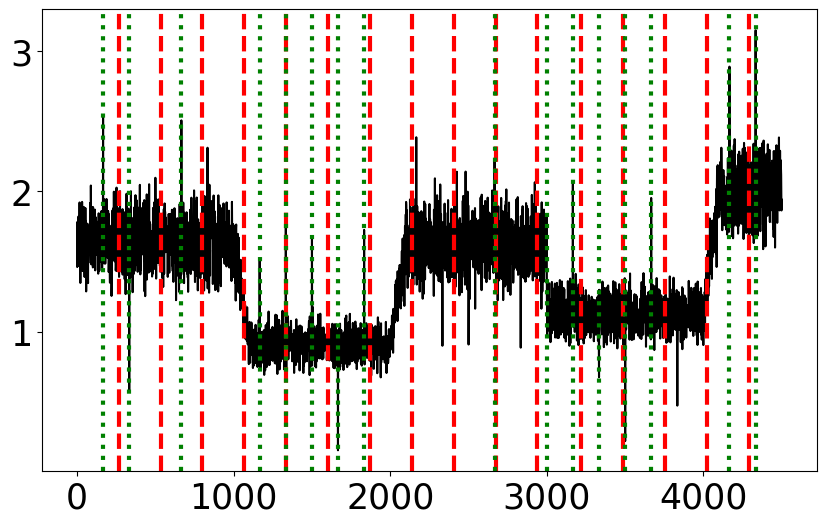}   
        \end{minipage}%
    }
    \subfigure[RANSAC+KSWIN] 
    {
        \begin{minipage}[t]{0.19\textwidth}
            \centering          
            \includegraphics[width=1\textwidth]{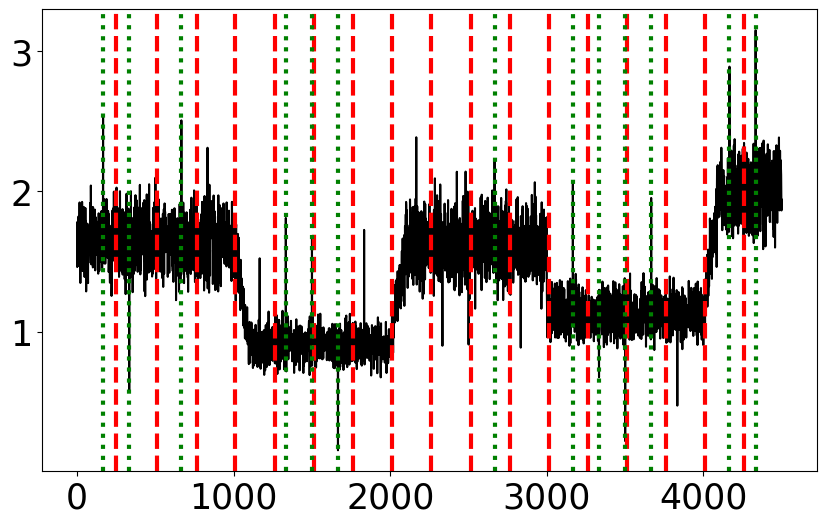}   
        \end{minipage}%
    }
    \subfigure[$\Theta$-IPOD+KSWIN] 
    {
        \begin{minipage}[t]{0.19\textwidth}
            \centering          
            \includegraphics[width=1\textwidth]{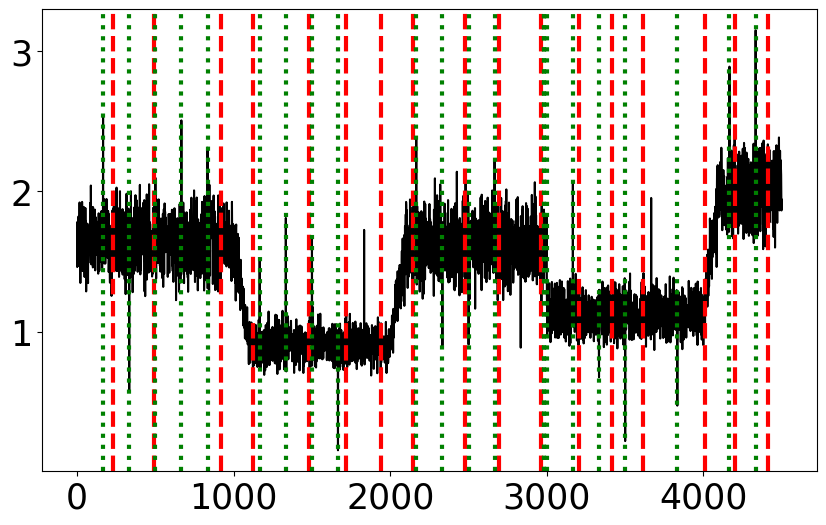}   
        \end{minipage}%
    }
    \subfigure[ARLF+KSWIN] 
    {
        \begin{minipage}[t]{0.19\textwidth}
            \centering          
            \includegraphics[width=1\textwidth]{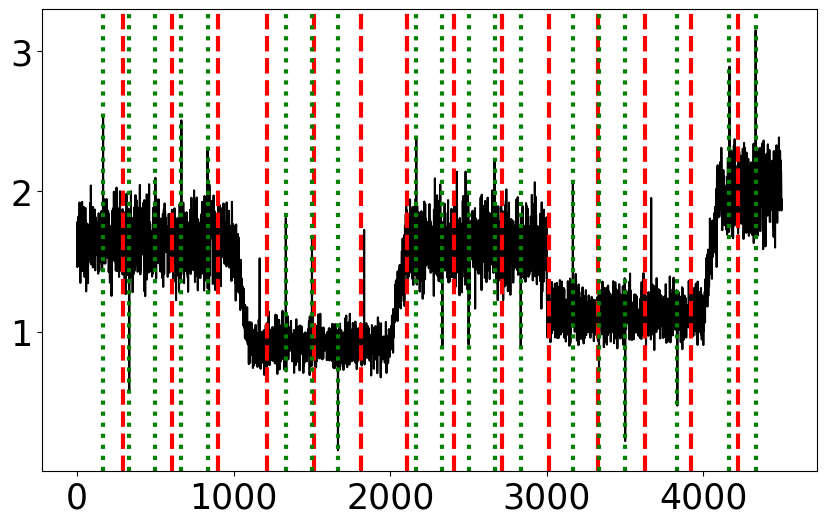}   
        \end{minipage}%
    }
    
    \subfigure[Huber+PH] 
    {
        \begin{minipage}[t]{0.19\textwidth}
            \centering          
            \includegraphics[width=1\textwidth]{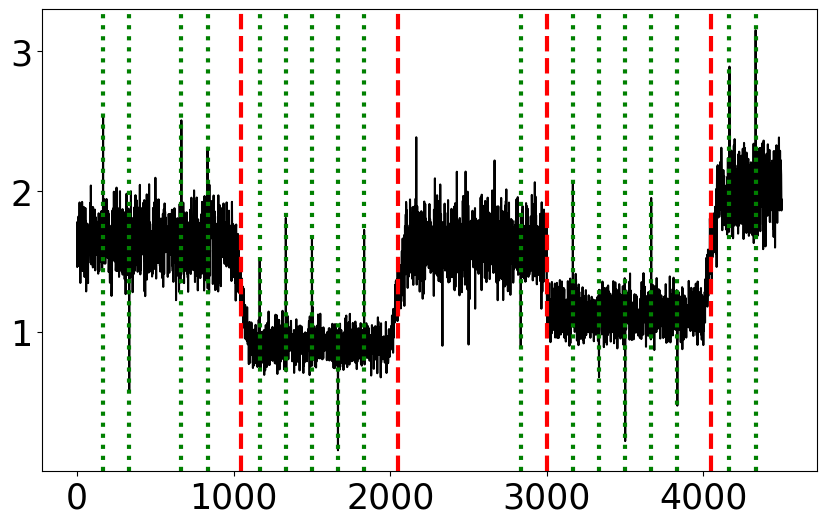}   
        \end{minipage}%
    }
    \subfigure[TheilSen+PH] 
    {
        \begin{minipage}[t]{0.19\textwidth}
            \centering          
            \includegraphics[width=1\textwidth]{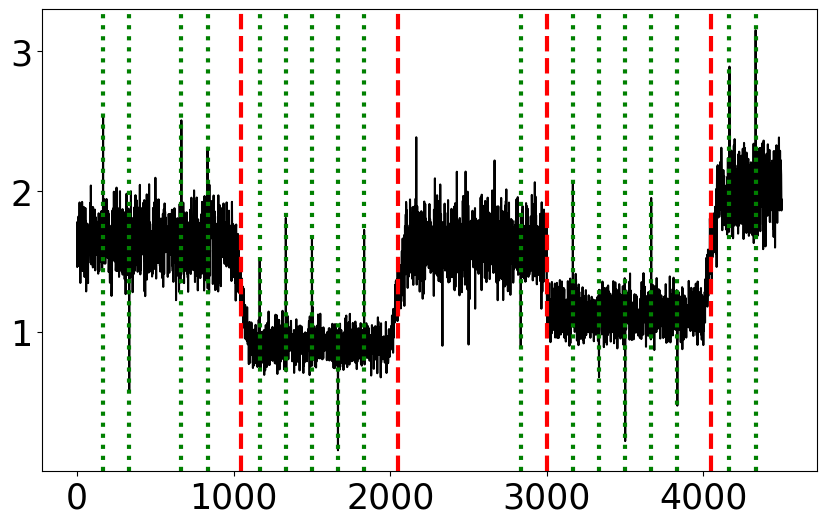}   
        \end{minipage}%
    }
    \subfigure[RANSAC+PH] 
    {
        \begin{minipage}[t]{0.19\textwidth}
            \centering          
            \includegraphics[width=1\textwidth]{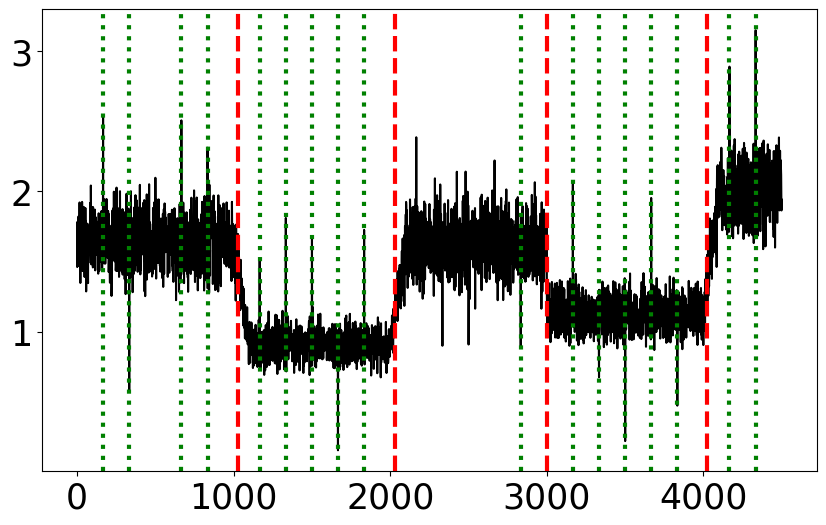}   
        \end{minipage}%
    }
    \subfigure[$\Theta$-IPOD+PH] 
    {
        \begin{minipage}[t]{0.19\textwidth}
            \centering          
            \includegraphics[width=1\textwidth]{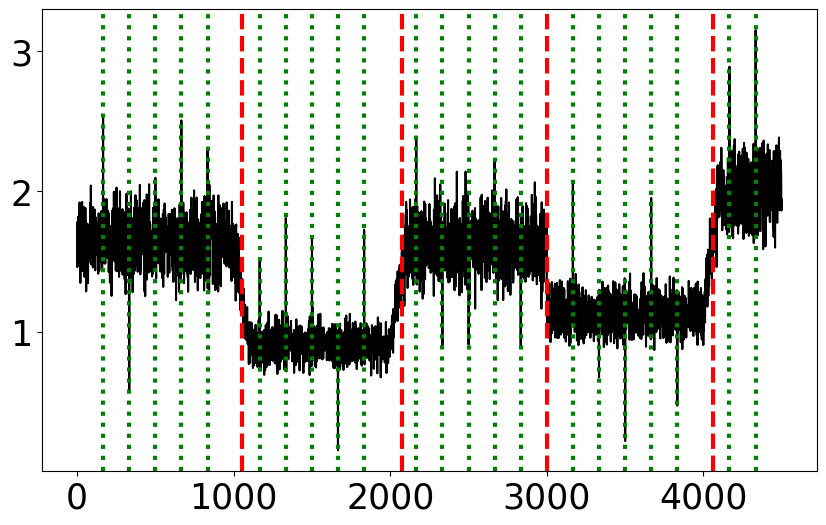}   
        \end{minipage}%
    }
    \subfigure[ARLF+PH] 
    {
        \begin{minipage}[t]{0.19\textwidth}
            \centering          
            \includegraphics[width=1\textwidth]{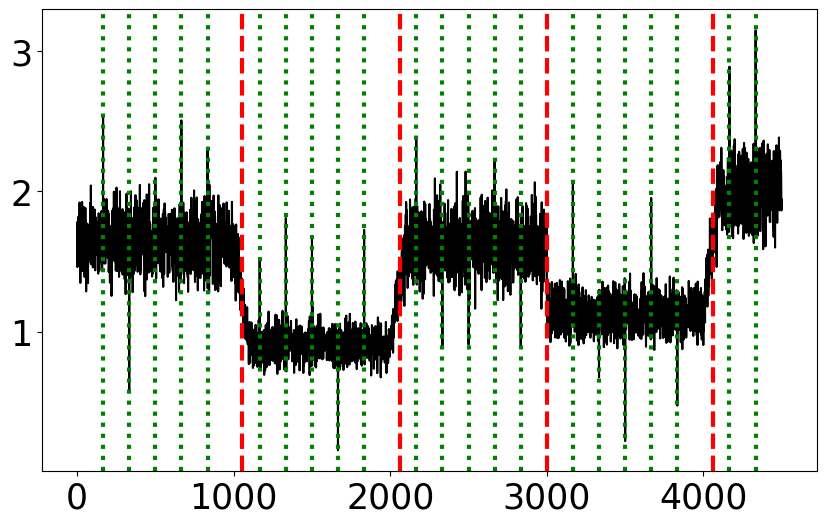}   
        \end{minipage}%
    }
    
    \subfigure[Huber+DSA] 
    {
        \begin{minipage}[t]{0.19\textwidth}
            \centering          
            \includegraphics[width=1\textwidth]{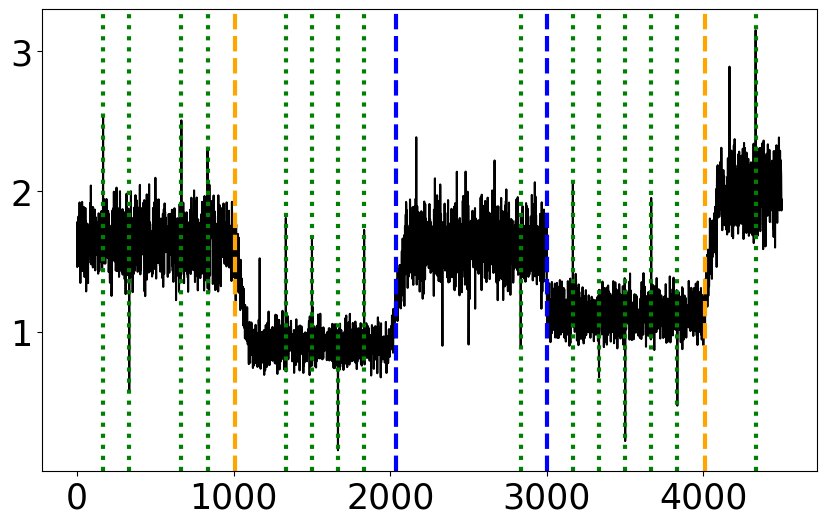}   
        \end{minipage}%
    }
    \subfigure[TheilSen+DSA] 
    {
        \begin{minipage}[t]{0.19\textwidth}
            \centering          
            \includegraphics[width=1\textwidth]{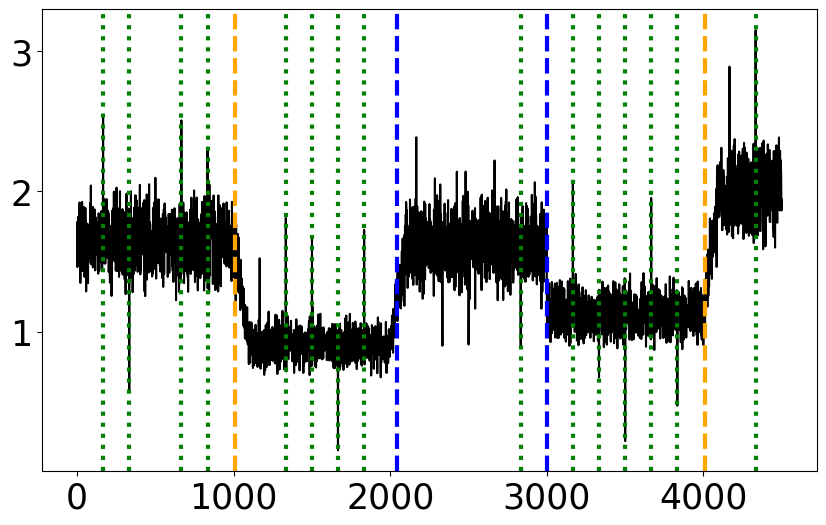}   
        \end{minipage}%
    }
    \subfigure[RANSAC+DSA] 
    {
        \begin{minipage}[t]{0.19\textwidth}
            \centering          
            \includegraphics[width=1\textwidth]{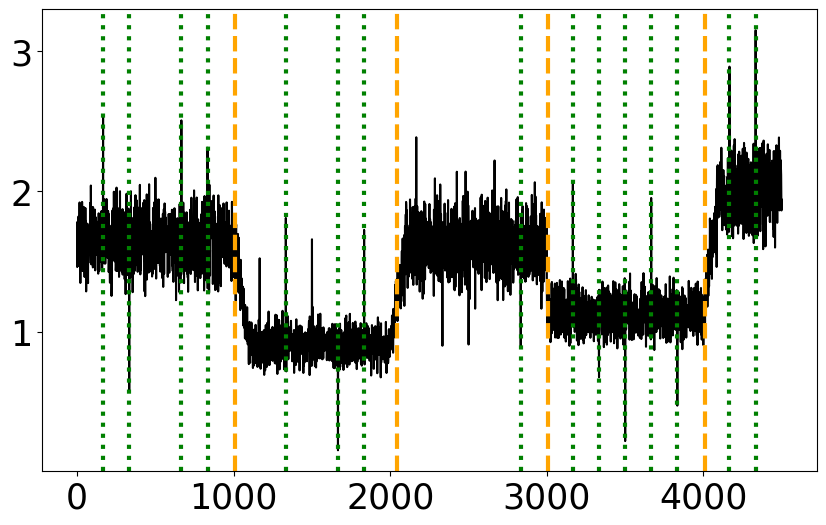}   
        \end{minipage}%
    }
    \subfigure[$\Theta$-IPOD+DSA] 
    {
        \begin{minipage}[t]{0.19\textwidth}
            \centering          
            \includegraphics[width=1\textwidth]{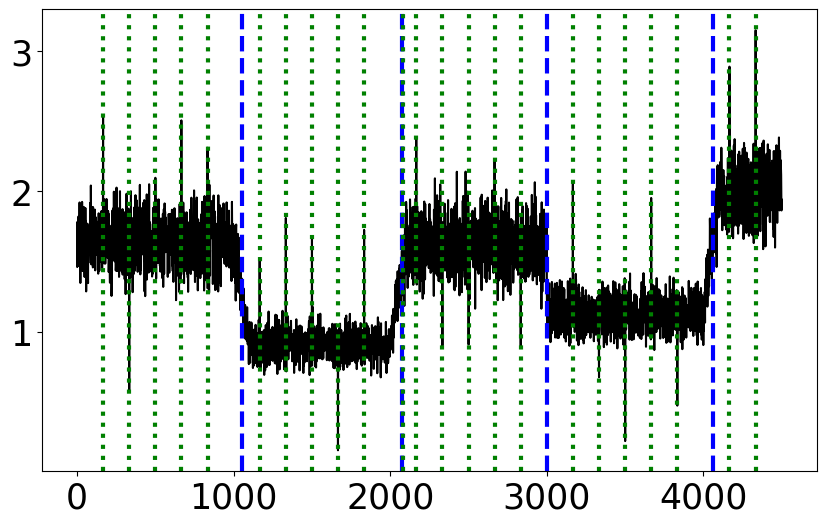}   
        \end{minipage}%
    }
    \subfigure[ARLF+DSA] 
    {
        \begin{minipage}[t]{0.19\textwidth}
            \centering          
            \includegraphics[width=1\textwidth]{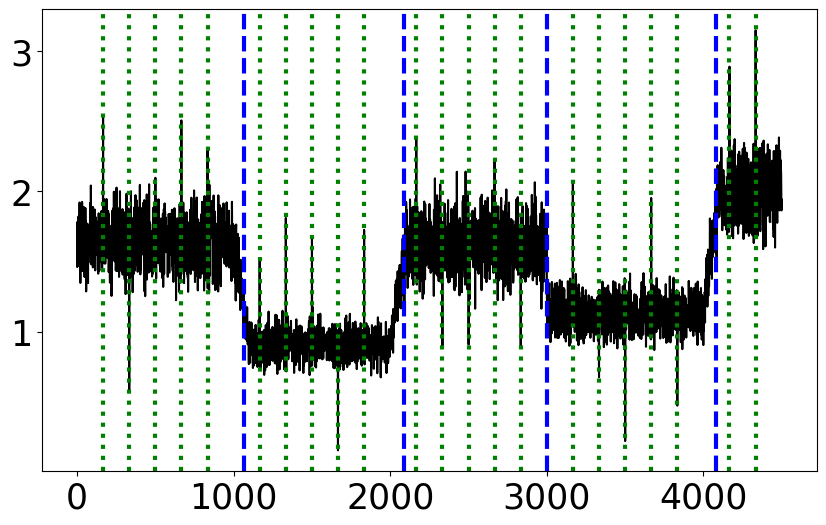}   
        \end{minipage}%
    }

    \subfigure[Huber+\colred{Ours}] 
    {
        \begin{minipage}[t]{0.19\textwidth}
            \centering          
            \includegraphics[width=1\textwidth]{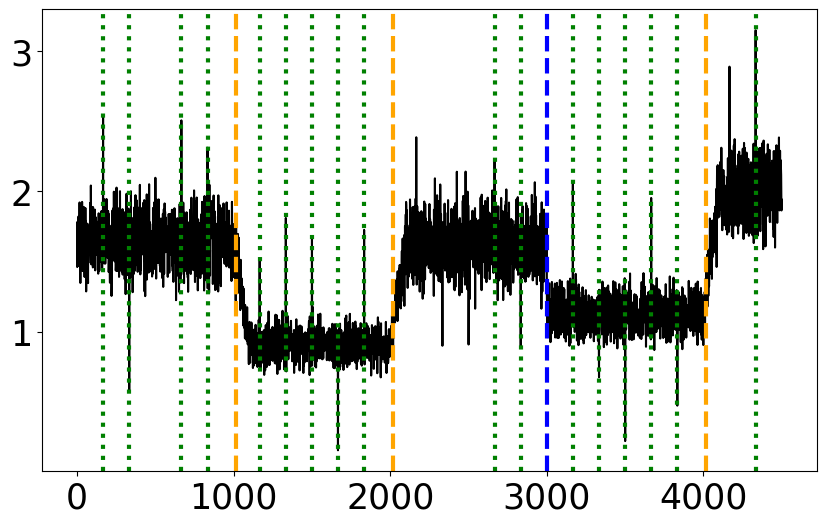}   
        \end{minipage}%
    }
    \subfigure[TheilSen+\colred{Ours}] 
    {
        \begin{minipage}[t]{0.19\textwidth}
            \centering          
            \includegraphics[width=1\textwidth]{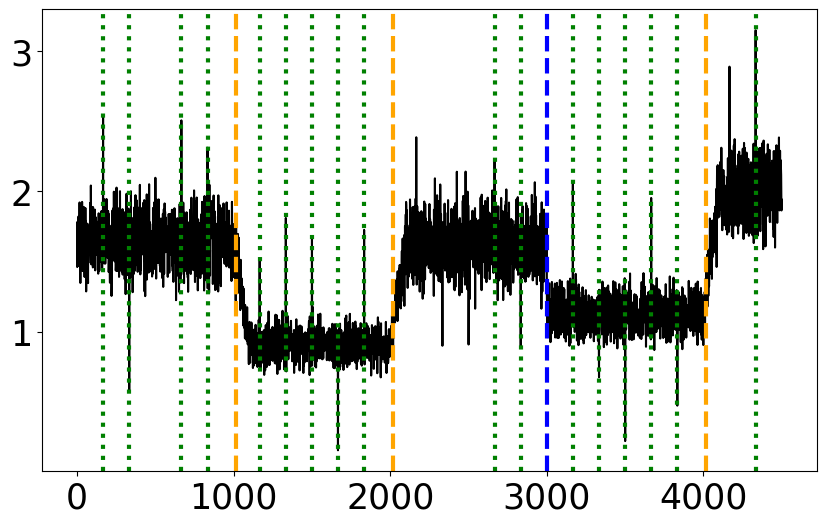}   
        \end{minipage}%
    }
    \subfigure[RANSAC+\colred{Ours}] 
    {
        \begin{minipage}[t]{0.19\textwidth}
            \centering          
            \includegraphics[width=1\textwidth]{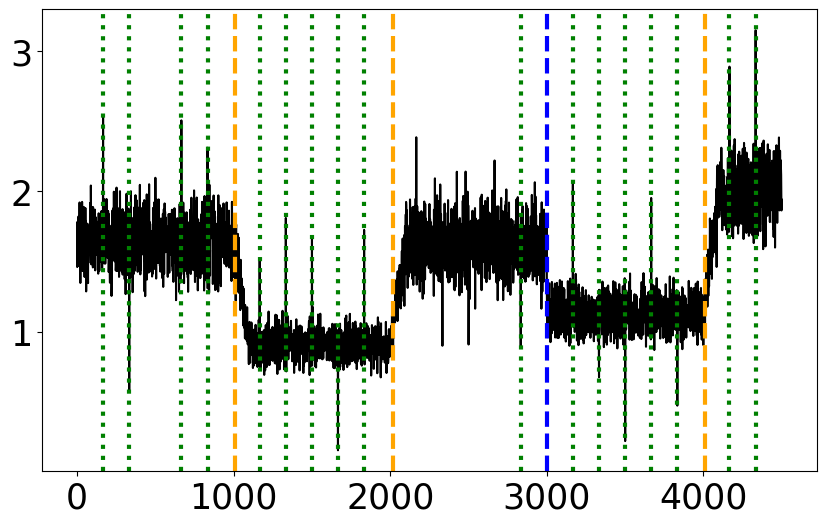}   
        \end{minipage}%
    }
    \subfigure[$\Theta$-IPOD+\colred{Ours}] 
    {
        \begin{minipage}[t]{0.19\textwidth}
            \centering          
            \includegraphics[width=1\textwidth]{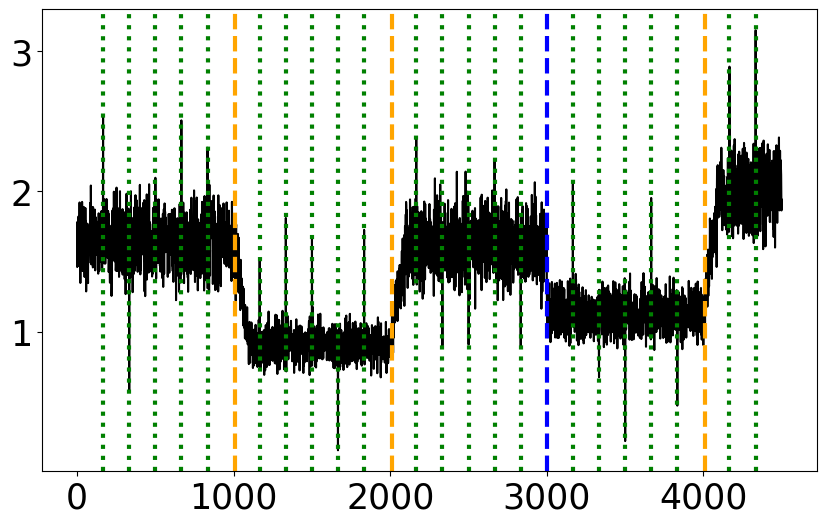}   
        \end{minipage}%
    }
    \subfigure[ARLF+\colred{Ours}] 
    {
        \begin{minipage}[t]{0.19\textwidth}
            \centering          
            \includegraphics[width=1\textwidth]{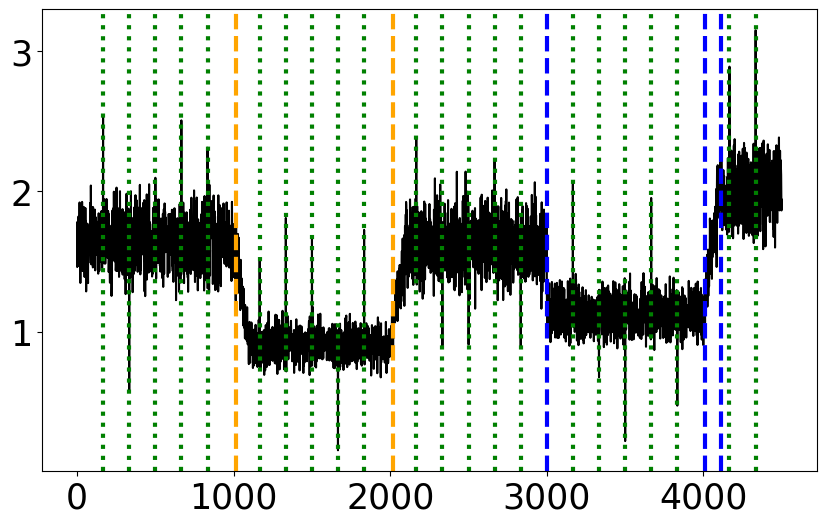}   
        \end{minipage}%
    }
    
    \footnotesize 
    \begin{tabular}{c c c c c}
        \legendline{black}{Stream}\quad
        \legendline{red, dashed}{Drift Detection} \quad 
        \legendline{green, dashed}{Outlier Detection} \quad
        \legendline{blue, dashed}{Abrupt Detection}\quad 
        \legendline{orange, dashed}{Incremental Detection} 
    \end{tabular}
    
    \caption{\small Performance of detection across different regression backbones and drift detectors on Mixed$_{0.005}$ dataset. } 
    \label{fig:vis}  
\end{figure*}

\subsection{Experimental results}
To effectively evaluate the performance of the proposed dual-channel decision process and EWMAD-DT detector, we set up the following four key research questions (RQs):

\textbf{RQ1}: Can the dual-channel decision process with EWMAD-DT effectively distinguish between true concept drifts and outliers?

\textbf{RQ2}: How responsive is the EWMAD-DT to distinct drift patterns, particularly in detection delay and accuracy for abrupt and incremental shifts?

\textbf{RQ3}: Does the superior detection capability of the proposed EWMAD-DT translate into lower regression errors (MAPE*) for robust regression tasks?

\textbf{RQ4}: Is the computational cost of EWMAD-DT acceptable for real-time stream processing given its performance gains?

\subsubsection{Answer to RQ1}
To answer RQ1, the proposed framework with EWMAD-DT is compared
 with four drift detectors on six datasets. 
 
\textbf{Mitigation of False Alarms against Outliers (The ``Defense'')}: As illustrated in Figs. \ref{fig:abr}, \ref{fig:inc} and \ref{fig:mix}, baseline methods exhibit a drop in drift detection precision when the outlier ratio increases. This empirical evidence confirms their susceptibility to the ``False Alarm'' problem, where high-frequency outliers are misinterpreted as abrupt drifts, triggering unnecessary and costly model resets. In stark contrast, our framework maintains a consistently higher Drift F1-score even under severe outlier contamination. This validates that Channel I (Outlier Detection) effectively acts as a robust filter, preventing transient noise from propagating to the drift detector and corrupting the state estimation. The success of this distinction stems from the structural decoupling in our dual-channel design. By handling outliers and drifts in separate logical flows, the system breaks the ambiguity inherent in regression residuals. Consequently, the EWMAD-DT detector in Channel II receives a ``clean'' signal, allowing it to focus exclusively on diagnosing the specific drift type.

\textbf{Superior Sensitivity to Different Drifts (The ``Offense'')}: Beyond noise suppression, results on the Mixed$_\delta$, Inc$_\delta$ and Abr$_\delta$ datasets reveal a fundamental limitation of general-purpose detectors. Traditional methods like ADWIN and KSWIN treat all residual deviations homogeneously, lacking the mechanism to distinguish the subtle, monotonic accumulation of residual from random fluctuations. Consequently, their detection accuracy is significantly lower than that of specialized methods like PH, DSA and our EWMAD-DT. By leveraging the type-differentiable mechanism of DSA and EWMAD-DT, our framework not only detects outlier, abrupt and incremental drifts but also accurately categorizes them. 

\textbf{Visualization of Detection Trajectory}:
To provide a qualitative assessment of the model's behavior in complex non-stationary environments, Fig. \ref{fig:vis} illustrates the detection trajectory on the Mixed$_{0.005}$ dataset. We visualize the distribution of the target variable $Y$ over time, as the evolution of $Y$ serves as the most direct proxy for concept drifts induced by changes in the underlying coefficient $\bm{\beta}_t$.
The window focuses on the first $4500$ instances, a period characterizing a highly volatile environment containing both incremental drifts (occurring at $t=1000, 2000, 4000$, $L=100$) and an abrupt drift (at $t=3000$). To rigorously evaluate detection timeliness. The sliding window size is fixed at $w=500$ to balance sensitivity and stability. As shown in the figure, the markers indicate the specific time points where the system triggered alarms for outliers, general drifts, abrupt drifts, or incremental drifts. While baseline methods often confuse drift types or exhibit significant latency, our framework demonstrates two distinct advantages: our framework accurately distinguishes between abrupt and incremental shifts, labeling them correctly in the visualization. The detection markers of our method align closely with the ground truth, confirming its superior detection precision and minimal adaptation lag.

\subsubsection{Answer to RQ2}
To address RQ2, we evaluate the responsiveness of EWMAD-DT by analyzing its detection delay and accuracy across distinct drift patterns. The results confirm that our method minimizes latency for abrupt shifts while maintaining high sensitivity to incremental changes, achieving a superior balance between delay and precision.

\textbf{Rapid Response to Abrupt Drifts}: As demonstrated in Figs. \ref{fig:abr}, our framework achieves the highest abrupt F1-score and lower mean detection delay when
the outlier ratio increases in abrupt drift scenarios. This rapid response is attributed to the exponential weighting mechanism of EWMAD-DT, which assigns higher significance to recent residuals. When an abrupt concept shift occurs, the sudden spike in prediction error is immediately amplified by the weighting factor, causing the monitoring statistic to cross the dynamic threshold almost instantaneously. This maximizes detection precision and minimizes the ``adaptation lag'', ensuring that the regression learner is reset promptly to capture the new concept. 

\textbf{Adaptive Tracking of Incremental Drifts}: For incremental drifts, which are inherently challenging to detect due to their slow accumulation, EWMAD-DT exhibits superior sensitivity. Unlike PH and DSA that require a large accumulation of error to trigger, EWMAD-DT employs a dynamic threshold that adjusts to the evolving volatility of the data stream, leveraging the sign of the $k$-th order difference to distinguish between various drift types. As shown in Figs. \ref{fig:inc} and \ref{fig:mix} (Detailed Drift Performance), our method achieves a significantly higher F1-score for incremental drifts and lower mean delay compared to DSA. This confirms that the dynamic thresholding effectively ``tightens'' the confidence interval as the model stabilizes, allowing it to detect subtle distributional shifts earlier than traditional methods.

\textbf{Limitation of Baselines}: As observed in Figs \ref{fig:abr}, \ref{fig:inc} and \ref{fig:mix}, while the PH test achieves a relatively high F1-score in low-noise scenarios, it exhibits two fundamental limitations. First, it treats all concept shifts homogeneously, lacking the mechanism to differentiate between abrupt and incremental drifts. Second, and more critically, its detection delay increases significantly as the outlier ratio increases. This indicates that PH struggles to accumulate sufficient statistical evidence in noisy environments, forcing it to ``wait longer'' to distinguish drifts from outliers, thereby sacrificing real-time responsiveness. In contrast, DSA and our EWMAD-DT demonstrate superior resilience. Both methods successfully distinguish between drift types. However, our EWMAD-DT stands out by maintaining the highest classification precision and consistently low detection delay for both Abrupt and Incremental drifts, even as the outlier ratio escalates.

\begin{table*}[H]
\centering
\caption{
Performance comparison of different concept drift detectors in terms of prediction accuracy (MAPE*) and computational efficiency (Time) across various robust regression learner.
}
\setlength{\tabcolsep}{0.9em} 
\scriptsize
\label{table:mape}
\resizebox{\textwidth}{!}{%
\begin{tabular}{cc cc cc cc cc cc} 
\toprule
\multirow{2}{*}{Dataset} & \multirow{2}{*}{Detector} 
& \multicolumn{2}{c}{ARLF} 
& \multicolumn{2}{c}{$\Theta$-IPOD} 
& \multicolumn{2}{c}{RANSAC} 
& \multicolumn{2}{c}{Huber} 
& \multicolumn{2}{c}{TheilSen} \\ 
\cmidrule(lr){3-4} \cmidrule(lr){5-6} \cmidrule(lr){7-8} \cmidrule(lr){9-10} \cmidrule(lr){11-12}
 & & MAPE* & Time & MAPE* & Time & MAPE* & Time & MAPE* & Time & MAPE* & Time \\ 
\midrule

\multirow{5}{*}{\shortstack{Inc$_{0.01}$}} 
& ADWIN & 0.0306 & 80.416 & 0.0246 & 25.195 & 0.0080 & 8.456 & 0.0080 & 8.074 & 0.0077 & 13.718 \\
& KSWIN & 0.0348 & 45.781 & 0.0291 & 25.866 & 0.0261 & 10.333 & 0.0233 & 8.602 & 0.0269 & 20.360 \\
& PH    & 0.0136 & 19.682 & 0.0089 & 6.590 & 0.0087 & 7.927 & 0.0092 & 5.253 & 0.0101 & 8.620 \\
& DSA   & 0.0128 & 21.862 & 0.0083 & 19.899 & 0.0083 & 10.006 & \textbf{0.0040} & 6.836 & 0.0099 & 10.434 \\
& Ours  & \textbf{0.0127} & 28.064 & \textbf{0.0063} & 18.848 & \textbf{0.0080} & 11.385 & 0.0067 & 7.063 & \textbf{0.0069} & 14.962 \\
\midrule

\multirow{5}{*}{\shortstack{Inc$_{0.02}$}} 
& ADWIN & 0.0369 & 74.536 & 0.0260 & 23.218 & \textbf{0.0096} & 10.693 & 0.0084 & 7.052 & 0.0108 & 11.539 \\
& KSWIN & 0.0368 & 53.401 & 0.0331 & 26.995 & 0.0270 & 10.423 & 0.0192 & 6.180 & 0.0260 & 19.131 \\
& PH    & \textbf{0.0184} & 15.125 & 0.0116 & 7.210 & 0.0154 & 7.115 & 0.0155 & 6.129 & 0.0250 & 9.121 \\
& DSA   & 0.0184 & 22.133 & 0.0095 & 18.209 & 0.0177 & 7.390 & \textbf{0.0068} & 5.803 & 0.0183 & 10.421 \\
& Ours  & 0.0186 & 29.536 & \textbf{0.0084} & 18.109 & 0.0144 & 10.299 & 0.0111 & 6.403 & \textbf{0.0108} & 14.974 \\
\midrule

\multirow{5}{*}{\shortstack{Abr$_{0.01}$}} 
& ADWIN & 0.0322 & 61.532 & 0.0214 & 35.304 & 0.0065 & 10.393 & 0.0067 & 10.267 & 0.0064 & 12.942 \\
& KSWIN & 0.0326 & 50.580 & 0.0287 & 41.206 & 0.0210 & 11.843 & 0.0192 & 8.172 & 0.0260 & 18.790 \\
& PH    & 0.0097 & 18.485 & 0.0051 & 9.497 & 0.0090 & 7.880 & 0.0076 & 5.367 & \textbf{0.0056} & 9.110 \\
& DSA   & \textbf{0.0087} & 20.491 & 0.0051 & 12.714 & 0.0087 & 7.427 & \textbf{0.0053} & 7.524 & 0.0079 & 14.504 \\
& Ours  & 0.0092 & 19.807 & \textbf{0.0051} & 11.683 & \textbf{0.0019} & 8.264 & 0.0273 & 5.460 & 0.0203 & 11.182 \\
\midrule

\multirow{5}{*}{\shortstack{Abr$_{0.02}$}} 
& ADWIN & 0.0347 & 62.337 & 0.0245 & 24.636 & \textbf{0.0097} & 11.868 & 0.0077 & 7.681 & 0.0084 & 14.502 \\
& KSWIN & 0.0408 & 45.049 & 0.0324 & 25.864 & 0.0225 & 8.037 & 0.0195 & 7.352 & 0.0265 & 20.852 \\
& PH    & 0.0148 & 17.421 & 0.0073 & 6.919 & 0.0115 & 7.875 & 0.0145 & 6.094 & 0.0179 & 10.319 \\
& DSA   & 0.0141 & 26.704 & 0.0072 & 10.547 & 0.0114 & 7.358 & 0.0108 & 7.821 & 0.0144 & 15.300 \\
& Ours  & \textbf{0.0130} & 20.782 & \textbf{0.0072} & 8.102 & 0.0204 & 9.824 & \textbf{0.0012} & 5.919 & \textbf{0.0012} & 9.572 \\
\midrule

\multirow{5}{*}{\shortstack{Mixed$_{0.01}$}} 
& ADWIN & 0.0311 & 65.855 & 0.0228 & 17.265 & 0.0082 & 10.090 & 0.0069 & 6.729 & \textbf{0.0081} & 10.736 \\
& KSWIN & 0.0320 & 58.043 & 0.0295 & 18.054 & 0.0243 & 7.951 & 0.0187 & 7.356 & 0.0227 & 18.204 \\
& PH    & 0.0126 & 17.851 & 0.0080 & 5.384 & 0.0070 & 6.239 & 0.0082 & 4.982 & 0.0089 & 8.343 \\
& DSA   & \textbf{0.0116} & 18.266 & 0.0079 & 7.143 & 0.0081 & 7.567 & 0.0074 & 6.056 & 0.0085 & 10.345 \\
& Ours  & 0.0118 & 24.800 & \textbf{0.0058} & 9.900 & \textbf{0.0064} & 7.062 & \textbf{0.0067} & 5.792 & 0.0088 & 10.472 \\
\midrule

\multirow{5}{*}{\shortstack{Mixed$_{0.02}$}} 
& ADWIN & 0.0376 & 91.406 & 0.0278 & 21.053 & \textbf{0.0083} & 11.994 & \textbf{0.0083} & 9.193 & 0.0096 & 14.253 \\
& KSWIN & 0.0350 & 63.376 & 0.0312 & 20.594 & 0.0238 & 7.903 & 0.0234 & 5.486 & 0.0237 & 14.236 \\
& PH    & 0.0160 & 19.314 & 0.0100 & 7.158 & 0.0120 & 6.585 & 0.0162 & 5.052 & 0.0233 & 9.797 \\
& DSA   & 0.0164 & 22.449 & 0.0092 & 9.996 & 0.0145 & 8.877 & 0.0154 & 7.736 & 0.0192 & 11.357 \\
& Ours  & \textbf{0.0159} & 31.653 & \textbf{0.0078} & 9.934 & 0.0142 & 7.334 & 0.0197 & 6.499 & \textbf{0.0026} & 11.137 \\

\bottomrule
\end{tabular}
}

\end{table*}

\subsubsection{Answer to RQ3}
To address RQ3, we investigate whether the superior detection metrics of EWMAD-DT translate into tangible improvements in regression fidelity. Specifically, we analyze the Mean Absolute Percentage Error on valid samples (MAPE*), which excludes outliers and drift transition periods, to evaluate how well the model captures the true underlying concepts.

As evidenced in Table. A, our framework consistently achieves the lowest MAPE* across different regression backbones. By leveraging the dual-channel mechanism to identify and filter out transient outliers before model updates, the regression learner operates on a ``purified'' data stream. This effectively prevents the ``Masking Effect'', where extreme values disproportionately skew the regression parameters. Although MAPE* is calculated on valid samples, a slow detector causes the model to learn from ``mixed'' concepts during the transition phase, leading to the contamination of the model with obsolete historical statistics. The superior detection capability of EWMAD-DT is not an isolated metric but a prerequisite for robust regression. It translates into lower MAPE* by shielding the learner from outliers and ensuring it always aligns with the current concept through drift adaptation.

\subsubsection{Answer to RQ4}
To address RQ4, we scrutinize the trade-off between detection accuracy and computational cost, as detailed in Table \ref{table:mape}. 

Observing the results, EWMAD-DT achieves a superior balance between efficiency and accuracy.
In computationally intensive scenarios, such as the $\Theta$-IPOD regression setting, our method demonstrates significant efficiency gains. For instance, on the Mixed$_{0.02}$ dataset, EWMAD-DT requires only $9.934$s, reducing the execution time compared to ADWIN, KSWIN and DSA. This efficiency stems from our optimized detection strategy which avoids the exhaustive window re-calculations typical of ADWIN-based approaches. 
Compared to lightweight heuristic methods like PH and DSA, EWMAD-DT incurs a marginal increase in computational time. This slight overhead is intrinsic to our method's advanced diagnostic capabilities: unlike PH, EWMAD-DT explicitly distinguishes between abrupt and incremental drifts; furthermore, it calculates the precise endpoint of incremental drifts, a step not performed by DSA. While these additional operations require more CPU cycles, they are crucial for preventing model overreaction to noise and ensuring precise drift detection. Given that EWMAD-DT consistently delivers the lowest error rates while maintaining runtime comparable to PH and DSA or significantly faster than ADWIN and KSWIN, we conclude that its computational cost is highly acceptable for real-time applications requiring high robustness.

\begin{figure*}
\centering
\scalebox{0.33}{
\includegraphics{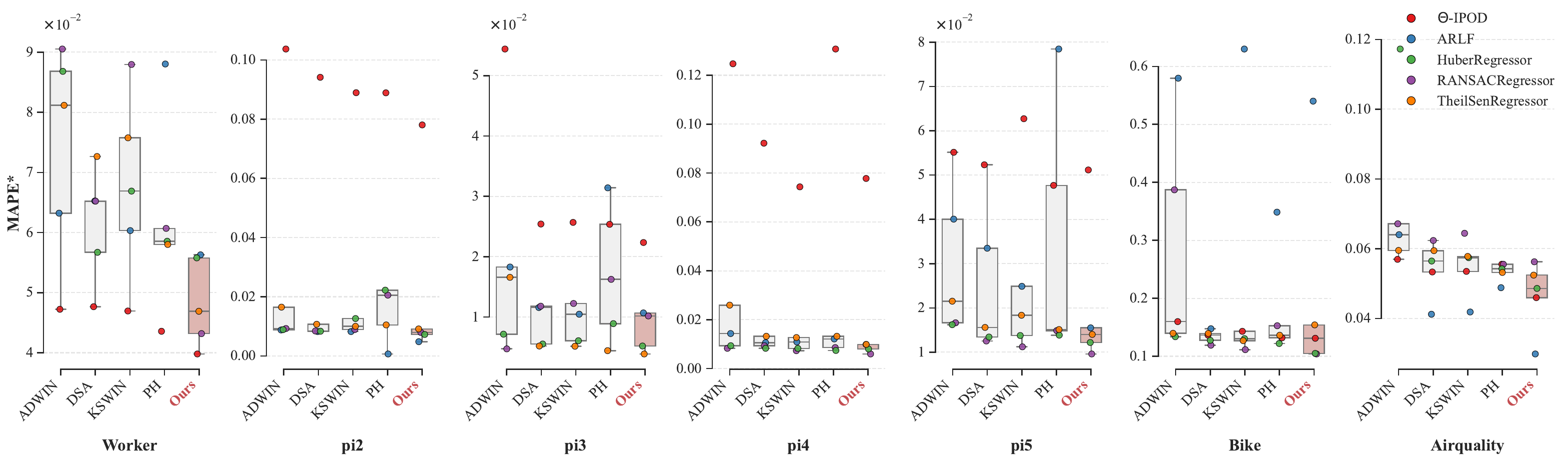}
}
\caption{Predictive precision across seven real-world datasets. We evaluate the MAPE* on clean samples (excluding outliers and drifts) across distinct domains. The proposed EWMAD-DT (Red) is benchmarked against other detectors (Grey). }
\label{fig:real_data}
\end{figure*}

\subsubsection{Comparison in real-world scenarios}
Fig. \ref{fig:real_data} illustrates the comparative performance of the proposed method against four drift detection baselines (ADWIN, DSA, KSWIN and PH) across seven real-world datasets. The evaluation metric is the Modified Mean Absolute Percentage Error on valid samples (MAPE*), where a lower value indicates superior predictive accuracy. Following parameter tuning via SeqUD, we report the MAPE* distribution of the performance. 
The results reveal two critical advantages of our approach:

As observed, our method consistently achieves the lowest median MAPE* across the majority of datasets. Specifically, on the Worker, pi2, pi4, pi5 and Airquality datasets, the proposed approach demonstrates a significant performance margin compared to the runner-up baselines. In the Worker scenario, which is characterized by high volatility and real-world noise, the baseline methods (such as ADWIN and KSWIN) exhibit MAPE* values fluctuating above $6 \times 10^{-2}$. In contrast, our method—benefiting from the optimal parameter configuration identified by SeqUD—suppresses the MAPE* to approximately $4.5 \times 10^{-2}$. This indicates that our drift detection mechanism can more accurately identify concept drifts, thereby enabling the downstream regressors to adjust promptly and minimize prediction errors.

A crucial observation lies in the compactness of the boxplots. Each box represents the performance variance when the drift detector is coupled with $5$ distinct regression models.
On datasets like pi3 and pi5, the baselines (especially PH and ADWIN) show elongated boxes and high interquartile ranges (IQR), indicating that their performance is highly sensitive to the choice of the downstream regressor.
Conversely, our method maintains a significantly tighter IQR. This model-agnostic stability implies that the drift detection signals provided by our method are robust and accurate enough to stabilize various regression strategies.

In summary, the comparative analysis confirms that the proposed method demonstrates the superior generalization capability across heterogeneous domains. The superior performance can be attributed to the dual-channel decision process's ability to filter the outliers and the rapid response of EWMAD-DT minimizes the 'adaptation lag'. This ensures that the model is promptly reset and retrained on relevant data, preventing the pollution of the learner with obsolete historical statistics.

\section{Conclusion}\label{section:Conclusion}
In this paper, we addressed the complex challenge of robust stream regression where outliers and concept drifts often co-occur. We proposed a novel model-agnostic joint detection framework featuring a dual-channel decision process. By effectively decoupling data cleaning from model updating, this framework serves as a universal solution that empowers various regression models to adapt to dynamic environments.
Crucially, empowered by the EWMAD-DT detector, our framework achieves precise, fine-grained discrimination among outliers, abrupt drifts, and incremental drifts, thereby effectively circumventing the masking effects. By isolating outliers in the first channel and leveraging dynamic thresholds to differentiate drift types in the second, our approach ensures that the model reacts specifically and appropriately to each type of change.
Extensive experiments on both artificial and real-world datasets validated the superiority of our method. The results demonstrate that our framework not only yields remarkable predictive performance but also maintains high detection accuracy in scenarios characterized by severe noise and mixed drift patterns.
Given the lightweight nature of the proposed framework, our future work will focus on deploying it on resource-constrained edge devices to enable efficient on-device learning.

\section*{Acknowledgement}
This work was supported in part by the National Natural Science Foundation of China under Grant 12471246.

\bibliographystyle{elsarticle-num}


\bibliography{myref}

\end{document}